\documentclass[11pt,a4paper,epsf,epsfig,psfrag]{article}
\usepackage{jheppub}
\usepackage{amsmath,epsfig}
\usepackage{subfigure}
\usepackage{amssymb,amsfonts}
\usepackage{latexsym}
\usepackage{epsfig}
\newbox\pippobox
\def\be{\begin{equation}}
\def\ee{\end{equation}}
\def\bea{\begin{eqnarray}}
\def\eea{\end{eqnarray}}
\def\del          {\partial}

\def\ee           {{\rm e}}

\newcommand{\beq}{\begin{equation}}
\newcommand{\eeq}{\end{equation}}
\newcommand{\beqa}{\begin{eqnarray}}
\newcommand{\eeqa}{\end{eqnarray}}
\newcommand{\beqar}{\begin{eqnarray*}}
\newcommand{\eeqar}{\end{eqnarray*}}

\renewcommand{\eqref}[1]{(\ref{#1})}

\def\dd{{\delta}}

\catcode`\@=12

\title{Some aspects of QGP phase in a hQCD model}
\author[a]{ Rong-Gen Cai}
\author[b]{Shankhadeep Chakrabortty,}
\author[a]{Song He,}
\author[a]{Li Li,}

\affiliation[a]{State Key Laboratory of Theoretical Physics,
Institute of Theoretical Physics, Chinese Academy of Science,
Beijing 100190, People's Republic of China }
\affiliation[b]{Institute of Physics, Bhubaneswar 751 005, India}

\emailAdd{cairg@itp.ac.cn}
\emailAdd{sankha@iopb.res.in}\emailAdd{hesong@itp.ac.cn}\emailAdd{liliphy@itp.ac.cn}

\date{\today}

\abstract{We continue to study the holographic QCD (hQCD) model,
proposed in a previous paper, in an Einstein-Maxwell-Dilaton (EMD)
system. In this paper we discuss some aspects of quark gluon plasma
(QGP) in the hQCD model, such as drag force, jet quenching parameter
and screening length. The results turn out to be consistent with
those as expected in QCD qualitatively. By calculating free energy
of the background black hole solution, we find that there exists a
 phase transition between small black hole and large black
hole when chemical potential $\mu $ is less than the critical one $
\mu_c$, and the phase transition is absent when chemical potential
is beyond the critical one.}


\begin{document}

\maketitle
\section{Introduction}
The analysis of various dynamical quantities from the experimental
data obtained at Relativistic Heavy Ion Collider (RHIC) in
Brookhaven National Laboratory leads to strong evidence for a strong
coupled quark gluon plasma (QGP), a deconfined phase of QCD at high
temperature and high number density \cite{RHIC:2012 eg}, \cite
{Shuryak:2008eq}, \cite{Shuryak:2004cy}. In the Au+Au collision with
the maximum center-of-mass energy around 200 GeV, several
phenomenological features, e.g, a very small value of shear
viscosity, quenching of high energy partons with large transverse
momentum, and elliptic flow etc., indicate that the dynamics of
thermal medium produced after collision is dominated by
non-perturbative effects~\cite{Baier:1996kr}. Being formulated in
Euclidean time, lattice QCD seems to be the best candidate to
explain the phenomena in thermal equilibrium, while the perturbative
QCD works only in the weak coupling region. But, both the
perturbative QCD and lattice QCD are failed to compute some
dynamical quantities like transport coefficients, drag force, and
jet quenching parameter etc., in the strong coupling regime. Thanks
to the feature of strong/weak coupling duality, the AdS/CFT
correspondence~\cite{dual} \cite{Aharony:1999ti} provides a powerful
tool to stress those issues.

In the AdS/CFT correspondence,  a well-known example of strong/weak
duality is AdS$_5$/CFT$_4$ and in this case, the four dimensional
conformal field theory is $\mathcal{N} = 4$ $SU(N_c)$ super
Yang-Mills (SYM), while the bulk theory is the type IIB supergravity
in AdS$_5$. In the AdS$_5$/CFT$_4$ correspondence, the ratio of
shear viscosity over entropy density for the dual field theory is
found to be small ($\eta/s=1/4\pi)$ for large number of colors
($N_C$) and large 't Hooft coupling ($\lambda = {g_{YM}}^2 N_C$)
\cite{Policastro:2001yc} \cite{Kovtun:2003wp} \cite{Buchel:2003tz}.
It turns out that this value is consistent with the one from RHIC
data with experimental error \cite{Teaney:2003kp}. In addition, it
is found that the value $1/4\pi$ is universal for various
deformations of the gauge theories with gravity duals.  Another
example of the universality is discussed in
\cite{Chakrabarti:2010xy}, where it is found that for some gauge
theories (not necessarily conformal) with well defined gravity duals
where if some conditions on the bulk stress energy tensor are
satisfied, the electrical conductivity at finite chemical potential
($\mu$) and temperature (T), the thermal conductivity ($\kappa_T$)
and the ratio of thermal conductivity to viscosity
($\frac{\kappa_T}{\eta T} \mu^2$) are independent of any specific
model.

Certainly a realistic holographic model dual to
 the strong coupled QCD at finite temperature and finite
density is a good starting point to study the RHIC physics.
Unfortunately such a model is still absent. However, due to the
universality mentioned above, one is expecting to understand various
features of QGP at RHIC with some deformed $AdS_5/CFT_4$, such as
drag force, jet-quenching parameter, and screening-length etc.
Indeed, over the past years a lot of works have been done along this
line.

By employing the so-called potential reconstruction approach
\cite{He:2010ye,Li:2011hp,He:2011hw,Cai:2012xh} to build up
holographic QCD model \cite{He:2010bx} from bottom up point of view,
 a hQCD model in an Einstein-Maxwell-dilaton theory is proposed in
\cite{Cai:2012xh}. In \cite{He:2011hw,Cai:2012xh},
 some properties of the hQCD model are studied, such
as equation of state, Wilson line operators \cite{He:2010ye} and
confinement/deconfinement phase transition and associated phase
diagram \cite{Cai:2012xh}, with good agreement with lattice QCD
simulation. It shows that this model can capture some most important
characteristics of realistic QCD. In this paper, we continue to
study the hQCD model by investigating some aspects of QGP phase such
as drag force, jet quenching parameter as well as screening length.
The results show that these quantities are consistent with QGP
properties. In this sense, we further confirm that this model may
give us the hints on the efficient way to deform the pure AdS$_5$
geometry and to realize the holographic description of low energy
QCD.

When high-energy partons perform a dragged motion as they pass
through the QGP medium, their energy loss can be encoded by drag
force. In the AdS/CFT correspondence, an external heavy quark and
its gluonic neighborhood are mapped into the endpoint of fundamental
string attached to the AdS boundary and the string itself in the AdS
bulk geometry, respectively. This external quark, with mass
proportional to the length of the string, loses its energy as the
string trailing back imposes a drag force on it. Within the
framework of gauge/gravity duality, the drag force experienced by an
external heavy quark moving with a constant velocity in $\mathcal{N}
= 4$ super Yang-Mills plasma at finite temperature is computed
\cite{Herzog:2006gh}, \cite{Gubser:2006bz}, \cite {Arnold:2007pg},
\cite{Sin:2006yz}, \cite{Herzog}, \cite{Friess:2006aw},
\cite{Casalderrey-Solana:2006rq}.  There are also further
generalizations of drag force computation for the charged
$\mathcal{N} = 4$ SYM \cite{Guijosa} and with the backreaction
effect due to the static heavy quark cloud distribution
\cite{Shankha:2010pm}.

Due to the medium, the suppression of heavy quark with high
transverse momentum leads to energy loss which is so called jet
quenching phenomenon \cite{Bjorken}, \cite{Adler:2005xv},
\cite{Bielcik:2005wu}. The transport coefficient $\hat{q}$
\cite{Baier:1996sk}, characterizing such phenomena, is defined
perturbatively as the ratio of square of the mean transverse
momentum over the mean free path \cite{D'Eramo:2010ak}. With the
framework of eikonal approximation \cite{Kovner:2003zj},
\cite{Kovner:2001vi}, the parameter $\hat{q}$ can be calculated from
light-like Wilson loop in adjoint representation \cite{Liu:2006ug},
\cite{Caceres:2006as}, \cite{Buchel:2006bv},
\cite{VazquezPoritz:2006ba}, \cite{Nakano:2006js},
\cite{Avramis:2006ip}, \cite{Gao:2006uf}. Ref.~\cite{Armesto:2006zv}
and \cite{Lin:2006au} calculate the jet quenching parameter in the
presence of chemical potential. Using AdS/CFT, the quark-antiquark
pair is mapped to the two endpoints of a fundamental string, both
endpoints attach on the AdS boundary. The string configuration shows
it hangs down to the bulk along radial direction and turns back to
the boundary again. The light-like Wilson loop in the fundamental
representation of boundary gauge theory and its thermal expectation
value correspond to the trajectory of two endpoints of the string
and $\exp(S)$ respectively, where $S$ is the string Nambu-Goto
action. In addition, we further extend our analysis by studying the
time-like Wilson loop and relate it with another important
parameter, the screening length of a quark-antiquark pair. It is
defined as the maximum length between a moving $q\bar{q}$ pair,
beyond which they break off with no binding energy and thus become
screened in the QGP medium. Screening length depends on the velocity
and the orientation of the quark-antiquark pair with respect to the
medium. Treating $\mathcal{N} = 4$ SYM at finite temperature as a
boundary theory, the binding energy between quark and antiquark pair
moving in the QGP and the screening length are calculated in
\cite{Liu:2006nn}. There are also other important generalizations to
calculate these quantities by introducing a boost on the original
static background \cite{Caceres:2006ta}, \cite{Chernicoff:2006hi},
\cite{Chernicoff:2008sa}. In this paper we will calculate the
screening length in the static frame of  $q\bar{q}$ pair for our
hQCD model.

In \cite{Cai:2012xh} we calculated heavy quark potential between a
quark-antiquark pair in our model, and found that there is a
confinement/deconfinement phase transition, and that there is a
critical point in the $T-\mu$ phase diagram. In this paper we
further confirm this phase transition in the small $\mu$ region by
computing free energy of the background black hole solution by using
the method in \cite{Kiritsis:2012ma}. In the AdS/CFT correspondence,
the Hawking-Page phase transition \cite{Hawking-Page} between AdS
black hole and thermal gas in AdS space is identified with the
confinement/deconfiement transition in gauge
theory~\cite{Witten:1998zw}\cite{hep-th/0608151}\cite{Cai:2007zw}\cite{Gursoy:2008za}.
In our case, the phase transition happens between small black hole
and large black hole, which will be clear shortly.  In addition, we
will argue that there is no phase transition between black hole
solution and thermal gas solution in our model.

The organization of the paper is as follows. In section 2 we briefly
review the potential reconstruction approach to the
Einstein-Maxwell-Dilaton system by generalizing the discussion in
\cite{Cai:2012xh} to the case with a coupling between dilaton field
and Maxwell field. In section 3, we discuss the generic black hole
solutions with asymptotical AdS boundary, and in particular present
an analytic black hole solution. In addition, in this section we
also briefly review the black hole solution for the hQCD model
studied in \cite{Cai:2012xh}. In section 4, we calculate the drag
force in this hQCD model. The jet quenching parameter and screening
length are discussed in section 5 and 6, respectively. In section 7,
we study the free energy of the background black hole solution and
discuss the phase transition between small black hole and large
black hole in small chemical potential region. Section 8 is devoted
to conclusions and discussions.

\section{Einstein-Maxwell-Dilaton  system}

\label{gravitysetup}
 In this section, we use the potential
reconstruction approach \cite{He:2010ye,Cai:2012xh} to study a 5D
Einstein-Maxwell-Dilaton (EMD) system. In \cite{Cai:2012xh}, the
authors did not consider the coupling between gauge field and
dilaton field in Einstein frame. Here we take the coupling into
consideration in a more generic version
\begin{equation} \label{minimal-String-action}
S_{5D}=\frac{1}{16 \pi G_5}\int d^5 x \sqrt{-g^S} e^{-2 \phi}
 \left(R^S + 4\partial_\mu \phi
\partial^\mu \phi-
V_S(\phi)-\frac{Z(\phi)}{4g_{g}^2}e^{\frac{-4\phi}{3}}F_{\mu\nu}F^{\mu\nu}\right),
\end{equation}
where the action (\ref{minimal-String-action}) is written in string
frame, $F_{\mu\nu}=\partial_\mu A_\nu-\partial_\nu A_\mu$ is the
Maxwell field, $Z(\phi)$ is an arbitrary function of dilaton field
$\phi$ and $V_S(\phi)$ is the dilaton potential. In Einstein frame
we can rewrite the action as \cite{He:2010ye}
\begin{equation} \label{minimal-Einstein-action}
S_{5D}=\frac{1}{16 \pi G_5} \int d^5 x
\sqrt{-g^E}\left(R-\frac{4}{3}\partial_{\mu}\phi\partial^{\mu}\phi-V_E(\phi)
-\frac{Z(\phi)}{4g_{g}^2}F_{\mu\nu}F^{\mu\nu}\right),
\end{equation}
where $ V_S=V_E e^{\frac{-4\phi}{3}}.$ The metrics in these two frames
are connected by the scaling transformation
\begin{equation}
g^S_{\mu\nu} = e^{4 \phi \over 3 }g^E_{\mu\nu}.
\end{equation}
The Einstein equations from the action
(\ref{minimal-Einstein-action}) read
\begin{eqnarray} \label{EOM}
E_{\mu\nu}+\frac{1}{2}g^E_{\mu\nu}\left(\frac{4}{3}
\partial_{\mu}\phi\partial^{\mu}\phi+V_E(\phi)\right)
-\frac{4}{3}\partial_{\mu}\phi\partial_{\nu}\phi -\frac{Z(\phi)}{2
g_{g}^2}\left(F_{\mu k}{F_\nu}^k-\frac{1}{4}
g^E_{\mu\nu}F_{kl}F^{kl}\right)=0,
\end{eqnarray}
where $E_{\mu\nu}=R_{\mu\nu}-\frac{1}{2}Rg_{\mu\nu}$ is Einstein
tensor. We here consider the ansatz $A=A_0(z) dt,\text{{ }} \text{{
}}\phi=\phi(z)$ for matter fields and
\begin{equation} \label{metric-stringframe}
ds_S^2 = \frac{{\ell^2} e^{2A_s}}{z^2}
\left(-f(z)dt^2+\frac{dz^2}{f(z)}+dx^{i}dx^{i}\right),
\end{equation}
for the metric in string frame, where $i=1,2,3$, $\ell$ is the radius of ${\rm
AdS}_5$ space, and $A_s$ is the warped factor, a function of
coordinate $z$. The metric in the string frame will be used to
calculate the loop operator below. In Einstein frame the metric
reads
\begin{eqnarray} \label{metric-Einsteinframe}
ds_E^2&= &\frac{{\ell^2} e^{2A_e}}{z^2}\left(-f(z)dt^2
+\frac{dz^2}{f(z)}+dx^{i}dx^{i}\right),\nonumber\\
&=& \frac{{\ell^2} e^{2A_s-\frac{4\phi}{3}}}{z^2}\left(-f(z)dt^2
+\frac{dz^2}{f(z)}+dx^{i}dx^{i}\right),
\end{eqnarray}
with $A_e=A_s-2\phi/3$.
In the metric (\ref{metric-Einsteinframe}), the $(t,t), (z,z)$ and
$(x_i, x_i)$ components of Einstein equations are respectively
\bea\label{Einsteiineq} &{}&b''(z)+\frac{b'(z) f'(z)}{2
f(z)}-\frac{b'(z)^2}{2 b(z)}+\frac{4}{9} b(z) \phi
'(z)^2+\frac{{A_0}'(z)^2 Z(\phi)}{6 g^2_g f(z)}+\frac{V_E(\phi)
b(z)^2}{3 f(z)}=0, \nonumber\\
&{}& \phi '(z)^2-\frac{9 b'(z) f'(z)}{8 b(z) f(z)}-\frac{9
b'(z)^2}{4 b(z)^2}-\frac{3 {A_0}'(z)^2 Z(\phi)}{8 g^2_g b(z)
f(z)}-\frac{3 V_E(\phi) b(z)}{4 f(z)}=0,
\nonumber\\
&{}&f''(z)+\frac{3 b'(z) f'(z)}{b(z)}+\frac{4}{3} f(z) \phi
'(z)^2+\frac{3 f(z) b''(z)}{b(z)}-\frac{3 f(z) b'(z)^2}{2
b(z)^2}-\frac{{A_0}'(z)^2 Z(\phi)}{2 g^2_g b(z)}+V_E(\phi) b(z)=0
,\nonumber \\
\eea
 where $b(z)=\ell^2 e^{2A_e}/z^2 $, and
$A_0(z)$ is electrical potential of Maxwell field. From those three
equations one can obtain following two equations which do not
contain the dilaton potential $V_E(\phi)$,
\begin{eqnarray}\label{AF}
 &{}&A_s''(z)+A_s'(z) \left(\frac{4 \phi '(z)}{3}+\frac{2}{z}\right)-A_s'(z)^2-\frac{2
 \phi ''(z)}{3}-\frac{4 \phi '(z)}{3 z}=0,\\\label{ff}
 &{}&f''(z)+ f'(z)\left(3 A_s'(z)-2 \phi '(z)-\frac{3 }{z}\right)-\frac{z^2 Z(\phi)
 e^{\frac{4 \phi (z)}{3}-2 A_s(z)} A_0'(z){}^2}{ g_{g}^2 L^2}=0.
\end{eqnarray}
Eq.(\ref{AF}) is our starting point to find exact solutions of the
system.  Note that Eq.(\ref{AF}) in the EMD system is the same as
the one in the Einstein-dilaton system considered in
\cite{Li:2011hp}\cite{He:2011hw} and the last term in Eq.(\ref{ff})
is an additional contribution from electrical field. In addition,
the equation of motion (EOM) of the dilaton field is given by
\begin{equation}
\label{fundilaton} \frac{8}{3} \partial_z
\left(\frac{\ell^3e^{3A_s(z)-2\phi} f(z)}{z^3}
\partial_z \phi\right)-
\frac{\ell^5e^{5A_s(z)-\frac{10}{3}\phi}}{z^5}\partial_\phi
V_E(\phi)+ \frac{Z'(\phi)b(z) A_0'(z)^2}{2 g_g^2}=0.
\end{equation}
And the EOM of the Maxwell field is given by \bea
\frac{1}{\sqrt{-g^E}}
\partial_\mu \left(\sqrt{-g^E}Z(\phi) F^{\mu\nu}\right)=0.\eea

From equations of motion, once $A_s(z)$ is given, we can obtain a
general solution to the system, which takes the following form \bea
\label{solutionU(1)1}\phi(z)&=&\int_0^z \frac{e^{2A_s(x)}
\left(\frac{3}{2} \int_0^x y^2 e^{-2 A_s(y)} A_s'(y)^2 \, dy+\phi
_1\right)}{x^2} \, dx+\frac{3 A_s(z)}{2}+\phi _0,\\
A_0(z)&=&A_{00}+A_{01} \left(\int_0^z {y e^{\frac{2 \phi
(y)}{3}-A_s(y)} \over Z(\phi(y))}\, dy\right),\\\label{f(z)}
f(z)&=&\int _0^zx^3 e^{2 \phi (x)-3 A_s(x)} \left(\frac{A_{01}{}^2
\left(\int_0^x  {y e^{\frac{2 \phi
(y)}{3}-A_s(y)}\over Z(\phi(y))}\, dy\right)}{ g_{g}^2 \ell^2}+f_1\right)dx+f_0,\\
V_E(z)&=&\frac{e^{-2 A_s(z)+\frac{4 \phi(z)}{3}} z^2 f(z)}{\ell^2}2
\Big(-\frac{e^{-2 A_s(z)+\frac{4 \phi(z)}{3}}Z(\phi(z)) z^2
A_0'(z)^2}{4 g_g^2\ell^2 f(z)}\nonumber\\&-&\frac{2 \left(3+3 z^2
A_s'(z)^2+4 z \phi'(z)+z^2 \phi'(z)^2-2 zA_s'(z) \left(3+2 z
\phi'(z)\right)\right)}{z^2}\nonumber\\&-&\frac{f'(z) \left(-3+3 z
A_s'(z)-2 z\phi'(z)\right)}{2 z f(z)}\Big),\label{solutionU(1)4}
 \eea
where  $\phi_0, A_{00}, A_{01}, f_0, f_1$ are all integration
constants and can be determined by suitable UV and IR boundary
conditions. When $Z(\phi)=1$, the general solution reduces to the
one given in \cite{Cai:2012xh}. Thus we have given a generic
formulism to generate a set of exact solutions of the EMD system
with a given $A_s(z)$.

\section{General asymptotical AdS black hole solutions}
\label{general solution}

Since here we are only interested in the black hole solutions with
asymptotic $AdS$ boundary, we impose the boundary condition $f(0)=1$
at the AdS boundary  $z= 0$, and require $\phi(z), f(z), A_0(z)$ to
be regular at black hole horizon $z_h$ and AdS boundary $z=0$. There
is an additional condition $A_0(z_h)=0$, which corresponds to the
physical requirement that $A_\mu A^\mu=g^{tt}A_0A_0$ must be finite
at $z=z_h$.

 We can parameterize the function $f(z)$ in
Eq.(\ref{solutionU(1)1}) as
 \bea\label{ffunction}
  f(z)&=&1+
\frac{A^2_{01} }{2 g_{g}^2 {\ell^2}}\frac{\int_0^z
g(x)\left(\int_0^{z_h}g(r)dr \int_r^x {g(y)^{\frac{1}{3}}dy\over
Z(\phi(y))}\right)dx}{\int_0^{z_h}g(x)dx} -\frac{\int_0^z g(x)dx
}{\int_0^{z_h}g(x)dx},
 \eea
 where $f_0=1$,
$f_1=-\frac{A^2_{01}}{4g_{g}^2
{\ell^2}}\frac{\int_0^{z_h}g(x)\int_0^x{g(y)^{\frac{1}{3}}\over
Z(\phi(y))}dy+1}{\int_0^{z_h}g(x)dx}$ and
 \bea \label{xyfunction}
g(x)&=&x^3 e^{2 \phi (x)-3 A_s(x)}. \eea We expand the gauge field
near the AdS boundary to relate the two integration constants to
chemical potential and charge density, respectively,
\bea A_0(z)&\sim &
A_{00}+A_{01}{e^{\frac{2\phi(y)}{3}-A_s(y)}\over
Z(\phi(y))}
z^2 + \cdots,\label{chemical} \eea with
 \bea A_{00} &=&\mu,\\
A_{01}&=& \frac{\mu}{\int_0^{z_h}y {e^{\frac{2\phi}{3}-A_s(y)}\over
Z(\phi(y))}dy}=\frac{\mu}{\int_0^{z_h} {g(y)^{\frac{1}{3}}\over
Z(\phi(y))}dy}. \eea

The temperature of the black hole can be determined through the
function $f(z)$ in (\ref{ffunction}) as   \bea\label{temp}
T=\frac{1}{4\pi}|f'(z)|_{z=z_h}=\left|{A^2_{01}\over 4\pi g_g^2
\ell^2}\frac{ g(z_h)\int_0^{z_h}g(r)dr\int_r^{z_h}{g^{1\over
3}(y)\over Z(\phi(y))}dy-g(z_h)}{\int_0^{z_h}g(x)dx}\right|. \eea
Following the standard Bekenstein-Hawking entropy formula, from the
geometry given in Eq.(\ref{metric-Einsteinframe}), we obtain the
black hole entropy density $S$ as
\begin{equation}
\label{entrpy} S={\frac{A_{area}}{4 G_5 V_3}=
\frac{\ell^3}{4G_5}\left(\frac{e^{A_s-\frac{2}{3}\phi}}{z}\right)^3}\Big|_{z_h},
\end{equation}
where $V_3$ is the volume of the black hole spatial directions
spanned by coordinates $x_i$ in (\ref{metric-Einsteinframe}).

\subsection{An analytical black hole solution}

\label{appendix-solution} 
\begin{figure}[h]
\centering
\includegraphics[width=10 cm]{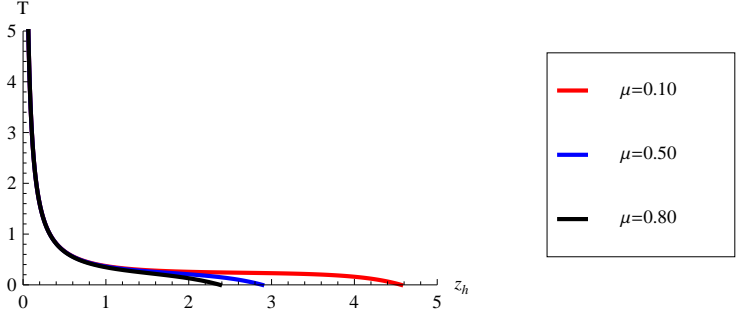}
\caption{The temperature as a function of horizon radius $z_h$ for
the analytical black hole solution with parameters $\ell=1,z_0=1,
g_g=1$.} \label{Sol1-T-zh}
\end{figure}

In this subsection, we list an analytical
 solution of the Einstein-Maxwell-Dilaton system by using
Eq.(\ref{solutionU(1)1}-\ref{solutionU(1)4}) with $Z(\phi)=1$. We
impose the constrain  $f(0)=1$, and require $\phi(z), f(z)$ to be
regular at $z=0$, and $z_h$. We give the solution in Einstein frame
as
\begin{equation} \label{Ametric-Einsteinframe}
ds_E^2 =\frac{{\ell^2} e^{2A_e}}{z^2}\left(-f(z)dt^2
+\frac{dz^2}{f(z)}+dx^{i}dx^{i}\right),
\end{equation}
with
\begin{eqnarray}
A_e(z)&=&\log \left(\frac{z }{z_0\sinh(\frac{z}{z_0})}\right),\nonumber\\
f(z)&=&1-\frac{4 V_{11}}{3}(3\sinh^4(\frac{z}{z_0})+2\sinh^6(\frac{z}{z_0}))+\frac{1}{8} V_{12}^2
 \sinh^4\left(\frac{z}{z_0}\right),\nonumber\\
\phi(z)&=&\frac{3 z}{2 z_0},\nonumber\\
A_0(z)&=& \mu -\frac{2g_g \ell}{z_0} V_{12}
\sinh^2\left(\frac{z}{2z_0}\right), \label{sol1}
\end{eqnarray}
where $z_0$ is an integration constant and $V_{11},V_{12}$ are two
constants from the dilaton potential \bea\label{dilatonpotential1}
V_E(\phi)&=&-\frac{12+9\sinh^2\left(\frac{2\phi}{3}\right)
+16V_{11}\sinh^6\left(\frac{\phi}{3}\right)}{{\ell^2}}+\frac{V_{12}^2
\sinh^6\left(\frac{2 \phi }{3}\right)}{8 {\ell^2}}. \eea The two
integration constants $V_{11}$ and $V_{12}$ then can be expressed in
terms of horizon $z_h$ and chemical potential $\mu$  as  \bea
V_{11}&=&\frac{3{\cosh}^4\left(\frac{z_h}{2 z_0}\right)
\left(\frac{\mu ^2 z_0^2 \sinh ^4\left(\frac{z_h}{z_0}\right)
{\cosh}^4\left(\frac{z_h}{2 z_0}\right)}{4 g_g^2
{\ell^2}}+8\right)}{32 \left(2 \sinh ^2\left(\frac{z_h}{2
z_0}\right)+3\right)},\nonumber\\V_{12}&=& \frac{\mu z_0
{\cosh}^2\left(\frac{z_h}{2 z_0}\right)}{2 g_g \ell}.
 \eea
 We can
obtain the temperature of the black hole by using Eq.(\ref{temp})
based on the above formulas. In Figure [\ref{Sol1-T-zh}], we show
the temperature as a function of horizon radius $z_h$ in cases of three different
 chemical potentials $\mu$. In this plot we take parameters $\ell=1,z_0=1, g_g=1$.
We see from Figure [\ref{Sol1-T-zh}] that the temperature with
respect to horizon $z_h$ is monotonic for a fixed chemical
potential. A vanishing temperature means that the black hole is
extremal with a smallest horizon radius. The smallest horizon radius
increases as chemical potential becomes large.

\subsection{The hQCD model}

Based on the general solutions, in Ref.~\cite{Cai:2012xh}, a
holographic QCD model is proposed to realize the
confinement/deconfinement phase transition of QCD. Since the aim of
this paper is to further study the hQCD model, we here briefly
review the main features of the model. In our hQCD model, the warped
factor $A_s(z)$ takes the form
\begin{equation} \label{ourmodel}
A_s(z)=  k^2 z^2,
\end{equation}
where $k$ is a constant.
To take this factor has various phenomenological motivations in order to build
a successful holographic QCD model, for details see \cite{Cai:2012xh}. In this paper,
we set $k=0.3\text{GeV}$, following Ref.~\cite{Cai:2012xh}. With this
factor, we  have the dilaton field $\phi$ as
\begin{eqnarray} \label{solu-phi}
\phi(z) &=& \frac{3}{4}  k^2 z^2(1+ H(z)),
\end{eqnarray}
where we have set the integration constant $\phi_0=0$, and $H(z)$ is given by
\begin{equation} \label{Hc}
H(z)=\, _2F_2\left(1,1;2,\frac{5}{2};2  k^2 z^2\right).
\end{equation} The characteristic function of the
black hole background takes the form
\begin{eqnarray}
 \label{solu-f}
f(z) = && 1+ \frac{1}{4 g_g^2
{\ell^2}}\left(\frac{\mu}{\int_0^{z_h}g(y)^{\frac{1}{3}}dy}\right)^2\frac{\int_0^z
g(x)\left(\int_0^{z_h}g(r)dr \int_r^x
g(y)^{\frac{1}{3}}dy\right)dx}{\int_0^{z_h}g(x)dx} \nonumber
\\
 && -\frac{\int_0^z g(x)dx }{\int_0^{z_h}g(x)dx},
\label{sol2}
\end{eqnarray}
where
\begin{eqnarray}\label{fc}
g(x)= x^3 e^{\frac{3}{2}  k^2 x^2(1+ H(x))-3k^2 x^2}.
\end{eqnarray}
One can clearly see  that the second term in (\ref{solu-f}) comes
from the contribution of electric field. If one turns off the
electric field, one can reproduce the black hole solution in
Einstein-dilaton system \cite{Li:2011hp}. In addition,
 the electric field $A_t(z)$ is given by
 \bea\label{A(z)}
A_t(z)=\mu + \frac{\mu}{\int_0^{z_h} g(y)^{\frac{1}{3}}dy} \int_0^z
x e^{\frac{1}{2}k^2 x^2(-1+H(x))}dx.
\eea

\begin{figure}[h]
\begin{center}
\epsfxsize=5.5 cm \epsfysize=5.5 cm \epsfbox{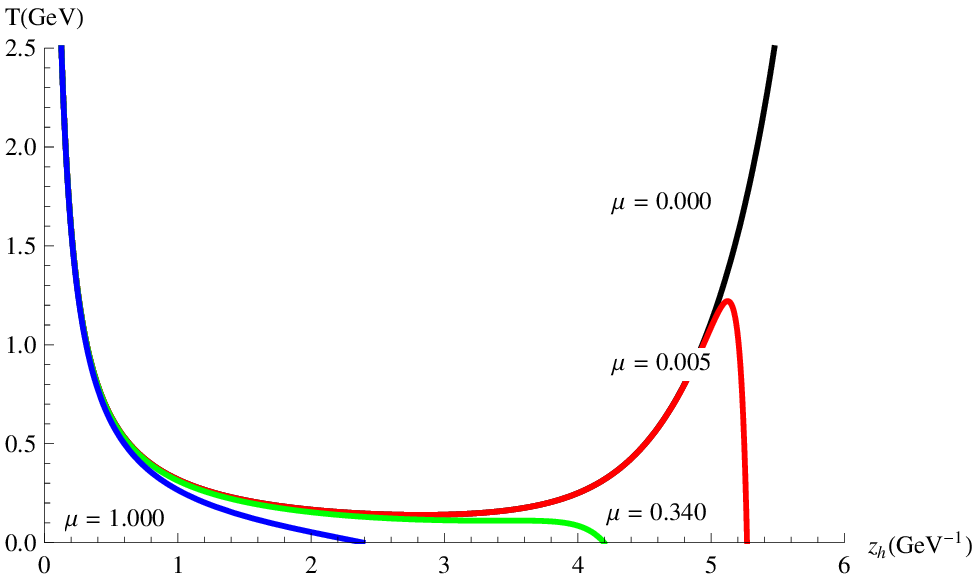}
\hspace*{0.1cm} \epsfxsize=5.5 cm \epsfysize=5.5 cm
\epsfbox{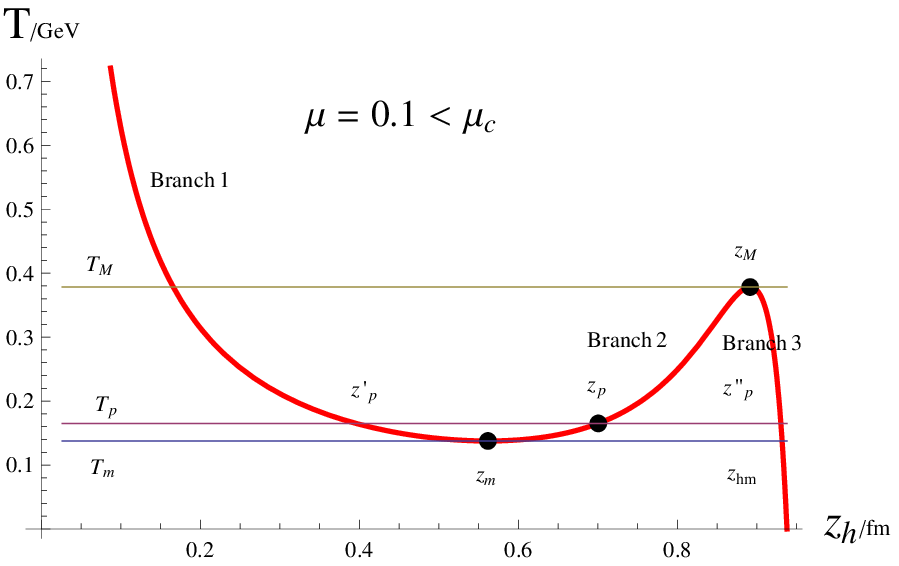} \vskip -0.05cm \hskip 0.15 cm
\textbf{( a ) } \hskip 6.5 cm \textbf{( b )} \\
\end{center}
\caption{Plot (a): The black hole temperature as a function of
horizon $z_h$ with different chemical potentials.  When $\mu>\mu_c$
the temperature  monotonically decreases to zero with increase of
$z_h$; when $0 <\mu <\mu_c$, the temperature decreases to a minimum
at $z_m$ and grows up to a maximum at $z_M$ and then decreases to
zero monotonically. When $\mu=\mu_c$, one has $z_m=z_M$.  Plot (b):
The temperature of the black hole with $\mu=0.1\text{GeV}$. The
three black hole solutions with horizon $z'_p$, $z_p$ and $z''_p$
have the same temperature. The black hole with $z_m <z_p<z_M$ is
thermodynamically unstable. Here we take $g_g \ell=1, k=0.3
\text{GeV}$. In this hQCD model, we always fix $ k=0.3 \text{GeV}$
and accordingly the critical chemical potential is $\mu_c=0.34
\text{GeV}$, which corresponds to the case $z_m=z_M$.} \label{T-zh}
\end{figure}

The temperature behavior of the black hole solution is discussed in
some details in \cite{Cai:2012xh}, with respect to horizon radius
and chemical potential. To be complete and for later use, we here
briefly repeat some main salient features. In Figure [\ref{T-zh}] we
plot the temperature with respect to horizon radius $z_h$ with
different chemical potentials. One can see clearly that the
temperature behavior crucially depends on the value of chemical
potential: there is a critical chemical potential $\mu_c$, beyond
which the black hole is always thermodynamically stable, while when
the chemical potential is less than the critical one, there is a
region of horizon radius, where the black hole is thermodynamically
unstable with negative heat capacity. To be more clear we plot in
Figure [\ref{T-zh}](b)  the temperature versus the horizon $z_h$ in
the case $\mu=0.1 {\rm GeV} <\mu_c$ as an example. One can see from
 the figure that the  black hole is
thermodynamically unstable in the region $z_m<z_h <z_{M}$ (branch 2
in plot b), where $z_m$ and $z_M$ are the black hole horizons
corresponding to the minimal and maximal temperatures ($T_m$ and
$T_M$), respectively. In this region, the heat capacity  of the
black hole is negative. The black hole solutions in the regions $z_h
<z_m$ and $z_h>z_M$ are thermodynamically stable (branch 1 and
branch 3). When $\mu\geq\mu_c$, $z_m$ and $z_M$ are degenerated to
one point. Note that in contrast to the case in Figure
[\ref{Sol1-T-zh}], there are local minimal and maximal values of
temperature in small $\mu$ cases. This is crucial to realize the
critical point in the $T-\mu$ phase diagram of the hQCD model
\cite{Cai:2012xh}.

 At this stage let us mention that it is quite interesting to
compare the temperature behavior of our black hole solution (see
Figure [\ref{T-zh}]) with the one for the Reissner-Nordstr\"om-AdS
black hole (see Figure 3 of Ref.~\cite{Chamblin:1999tk}). They look
quite similar. But there exist essentially different points between
them. At first, in our case, the temperature is plotted with respect
to horizon radius $z_h$ for a fixed chemical potential, while in
\cite{Chamblin:1999tk} it is plotted with respect to horizon radius
for a fixed charge (note that there the horizon radius $r_+$
corresponds to the inverse of horizon radius $z_h$ here). Second,
our black hole solution has a Ricci flat horizon, while the black
hole horizon discussed in \cite{Chamblin:1999tk} is a sphere. The
third is that we will discuss the phase transition in grand
canonical ensemble (for a fixed chemical potential) while the phase
transition between a small black hole and a large black hole
discussed in \cite{Chamblin:1999tk} is in canonical ensemble (for a
fixed charge). As we will see shortly that just due to the
similarity of temperature behavior, in our case, the phase
transition will also appear between a small black hole and a large
black hole in grand canonical ensemble when the chemical potential
is less than the critical value.

\section{Drag force}

In this section, mainly following \cite{Gubser:2006bz} we compute
the drag force experienced by an external probe quark traversing in
the QGP in the present hQCD model. The holographic computation of
the drag force is based on the gauge/gravity correspondence between
the deconfined phase of hQCD model at finite temperature and its
dual garvity realized as an aAdS black hole solution in the
Einstein-Maxwell-Dilaton background. In this aAdS black hole
geometry, the gravity dual of the probe quark is described by an
infinitely long fundamental string. One of its ends is attached to
the boundary of the bulk spacetime. The body of the string extends
along the radial direction and the free end of the string goes
parallel to black hole horizon. The gauge/gravity duality suggests
an identification between the end point of the string and the probe
quark. Furthermore, the body of
 the string captures the effects of thermal plasma through which the external quark is moving.
In this dual gravity picture, the string trails back and imparts a
drag force on it's endpoint that is attached to the boundary. This
drag force is obtained by calculating the rate of change in string
momentum.  The drag force is a function of temperature and chemical
potential. The boundary gauge theory we are considering is on
$\mathcal{M}^4$ described by the boundary coordinates $t, x^1, x^2,
x^3$. The dynamics of a fundamental string is completely specified
by the Nambu-Goto action in the black hole background within the
string frame (\ref{metric-stringframe}). On this background the
world sheet action reads
\begin{equation}
 S = - \frac{1}{2 \pi \alpha'} \int d\tau d\sigma \sqrt{-{\rm det}g_{\alpha \beta}}, \hspace{.5 cm} g_{\alpha\beta}
 = \frac{\del X^{\mu}}{\del \sigma_{\alpha}}
\frac{\del X^{\nu}}{\del \sigma_{\beta}}G_{\mu\nu},
\label{nambu-Goto}
\end{equation}
where $g_{\alpha \beta}$ is the induced metric on the world sheet and $G_{\mu \nu}$ is the background metric.
 The equation of motion derived from
 (\ref{nambu-Goto}) is given by
\begin{equation}
 \Delta_{\alpha} P^{\alpha}_{\mu} = 0,  \hspace{.5 cm} P^{\alpha}_{\mu} =  - \frac{1}{2 \pi \alpha'} G_{\mu \nu} \del^{\alpha}X^{\nu},
\end{equation}
where $\Delta_{\alpha}$ is the covariant derivative with respect to $g_{\alpha \beta}$ and $P^{\alpha}_{\mu}$ is
the world sheet current of
space time energy-momentum of the test string.
 We consider the motion of the string along $x^1$. In the gauge, $\tau = t$
and $\sigma = z$, the string dynamics can be completely specified by
the function $x^1(t,z)$. In this case, the Lagrangian reads
\begin{equation}
 \mathcal{L} = -\frac{1}{2 \pi \alpha'} \sqrt{\frac{1}{H} + \frac{f(z) (\del_z x^1)^2}{H}
 - \frac{(\del_tx^1)^2}{H f(z)}},
\label{lag}
\end{equation}
where $H$ is defined as
\begin{equation}
 H = \sqrt{\frac{z^2}{{\ell^2} e^{2A_s}}}.
\end{equation}
To capture the dragged motion of the quark in the boundary theory we
assume the following ansatz in the bulk~\cite{Gubser:2006bz}
\begin{equation}
 x^1(t,z) = vt + \xi(z).
\label{ansatz}
\end{equation}
Here we have assumed only the late time behavior of the string
motion. With this ansatz the  Lagrangian reduces to
\begin{equation}
  \mathcal{L} = -\frac{1}{2 \pi \alpha'} \sqrt{\frac{1}{H} + \frac{f(z) (\del_z \xi(z))^2}{H} - \frac{v^2}{H
  f(z)}}.
\label{lag1}
\end{equation}
The momentum which conjugates to $\xi(z)$ reads
\begin{equation}
 \Pi_{\xi} = \frac{\del \mathcal{L}}{\del \xi^{'}} = -\frac{\xi'}{2\pi \alpha'} \frac{f}{H}\sqrt{\frac{Hf}{f - v^2 + f^2
 {\xi^{'}}^2}}.
\label{mom}
\end{equation}
For the sake of consistency it is important to invert the equation
(\ref{mom}) and write it in the following way
\begin{equation}
 {\xi^{'}} = \sqrt{ \frac{{\Pi_{\xi}}^2 (f - v^2)}{\frac{f^2}{H^2}[\frac{1}{4\pi^2{\alpha^{'}}^2}Hf - {\Pi_\xi}^2
 H^2]}}.
\label{prof}
\end{equation}
Here the positive sign is taken due to the trailing nature of the
string profile \cite{Gubser:2006bz}. To obtain the string profile we
have to solve the differential equation (\ref{prof}). To have a real
$\xi (z)$, we further impose the constraints
\begin{eqnarray}
f(z) |_{z=z_v} &=& v^2, \nonumber \\
{\Pi_{\xi}}^2|_{z=z_v} &=& \frac{1}{4\pi^2 {\alpha^{'}}^2}
\frac{v^2}{H}, \label{constraint}
\end{eqnarray}
 so that one has $\xi'|_{z=z_v}=v^2/f^2$, keeping finite. The
profile of the string is defined in the region with $z <z_v $, That
is, there is a maximal value $z_v <z_h $ for the string profile.

The constraints are very useful to figure out the final form of the
drag force. Before to compute the drag force, we here mention the
relation between the drag force in the boundary field theory
 and the dissipation of momentum flowing down the
string, in light of AdS/CFT correspondence. In the boundary
theory the presence of the thermal medium results into dissipation
of energy and momentum of external quark until it reaches thermal
equilibrium with the medium. In the  bulk theory the momentum is
flowing down the string from the boundary to the bulk and the change
of momentum at a given spatial point on the world sheet for a given
time interval can be calculated. The identifications of the endpoint of the string attached to the boundary with the quark and of the string in the bulk with the thermal medium around the
quark suggest that the drag force can be realized in terms of the force
imparted by the string on its boundary endpoint. To calculate the
 change of string momentum due to its motion along $x_1$
direction, we consider a closed curve on the world sheet and study
how the momentum is conserved around this curve \cite{Fi: 2012de}.
\begin{figure}[h]
\centering
\includegraphics[width=7 cm]{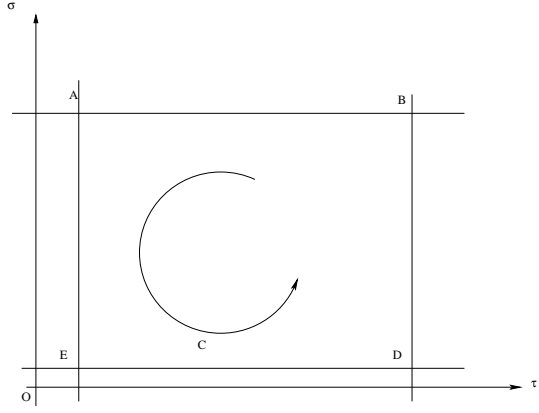}
\caption{This plot shows a closed path in an anti-clockwise
direction on a world sheet bounded by  coordinates $[A =
\tau_1,\sigma_2]$, $[ B = \tau_2,\sigma_2]$, $[ D =
\tau_2,\sigma_1]$ and $[E = \tau_1,\sigma_1]$.} \label{fig12}
\end{figure}
According to the conservation of  world sheet current of space time
energy-momentum of the test string, the total flux calculated around
the path $C$ must be zero,
\begin{equation}
 \oint_{ABDEA} (P^{\tau}_{\mu} d\sigma - P^{\sigma}_{\mu} d\tau) =
 0.
\label{cons1}
\end{equation}
Note that one end of the string is attached to the boundary and the
other end close to horizon is free. In the static gauge,
Eq.~(\ref{cons1}) reduces to
\begin{equation}
 p^{t_1}_{x_{1}} - p^{t_2}_{x_{1}} = - \int^{t_1}_{t_2} \sqrt{-g} P^{z}_{x_1} dt,
\end{equation}
where $p^{t}_{x_{1}}$ is the $x_1$ component of the total momentum
at time $t$. Consequently the drag force is defined as
\begin{equation}
 F_{drag} = \frac{dp_{x_1}}{dt} = - \sqrt{-g} P^{z}_{x_1} =
 -\frac{1}{2\pi\alpha^{'}} \frac{{\ell^2} e^{2A_s}}{{z_v}^2}
 v.
\label{drag}
\end{equation}
Finally we have to replace all the gravity parameters in terms of
gauge theory parameters. Before doing that, we analyze the form of
the constraints case by case. The exact forms of constraint for the
solutions (\ref{sol1}) and (\ref{sol2}) are given respectively by
\begin{eqnarray}
&\hspace{1 cm}v^2 = 1-\frac{4
V_{11}}{3}(3\sinh^4(\frac{z_v}{z_0})+2\sinh^6(\frac{z_v}{z_0}))+\frac{1}{8}
V_{12}^2
\sinh^4\left(\frac{z_v}{z_0}\right) ,  \nonumber \\
\hspace{-.3 cm} v^2 & =  1 + \frac{1}{4 g_g^2
{\ell^2}}\left(\frac{\mu}{\int_0^{z_h}g(y)^{\frac{1}{3}}dy}\right)^2\frac{\int_0^{z_v}
g(x)\left(\int_0^{z_h}g(r)dr \int_r^x
 g(y)^{\frac{1}{3}}dy\right)dx}{\int_0^{z_h}g(x)dx} -
\frac{\int_0^{z_v} g(x)dx }{\int_0^{z_h}g(x)dx}.
\label{con1}
\end{eqnarray}
It is always desirable to express the drag force in closed analytic
form as a function of gauge theoretical variables. However it is
very difficult to obtain  analytic forms for the constraint
 (\ref{con1}) and  the temperature (\ref{temp}).
Instead  we here solve them numerically and plot the  drag force
with respect to gauge theory parameters, e.g, temperature and
chemical potential, so that the qualitative features of the drag
 force can be revealed.

\begin{figure}
\centering {\includegraphics[width= 10 cm]{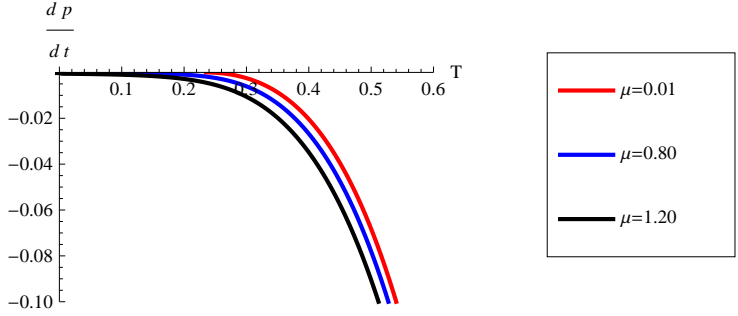}
\label{drag-T-mu ana}} \caption{This plot shows the drag force from
the analytic black hole solution as a function of $T$ for chemical
potential $\mu = 0.01 , 0.80$, and $ 1.20$ respectively. Here we
take $v=0.1$.}

\end{figure}

\begin{figure}
\centering \mbox{\subfigure[]{\includegraphics[width=5
cm]{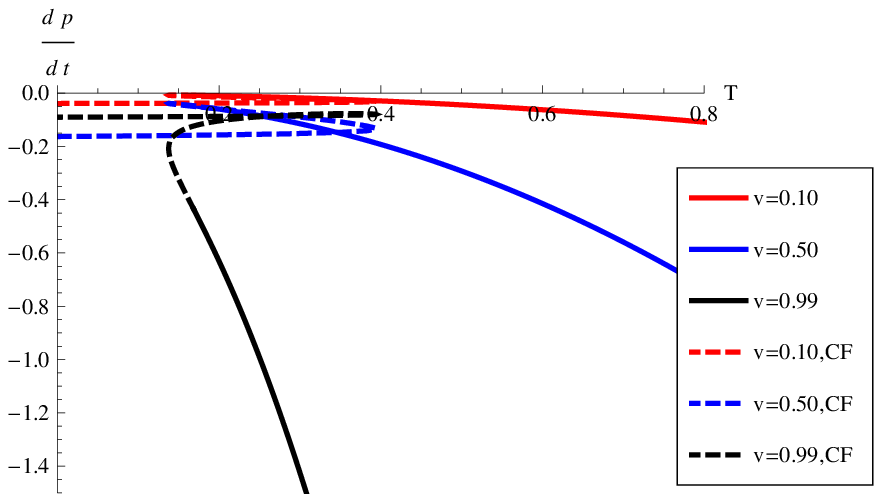}} \quad
\subfigure[]{\includegraphics[width=5
cm]{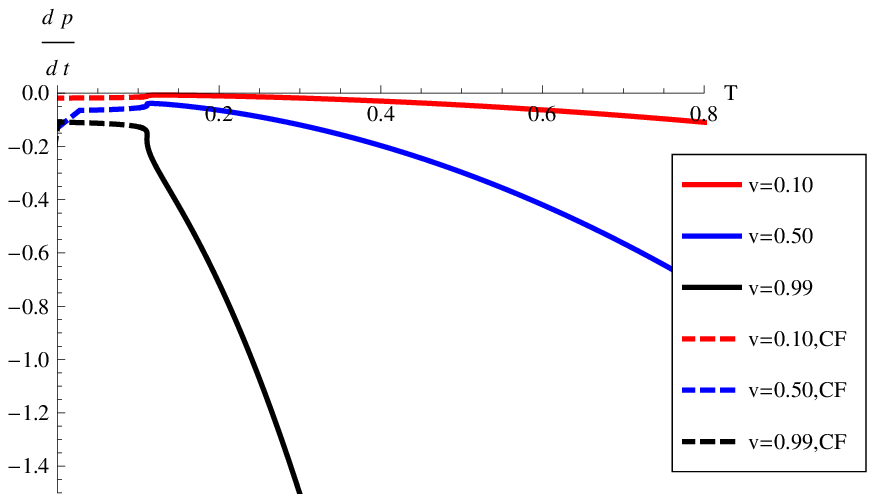} }
\subfigure[]{\includegraphics[width=5
cm]{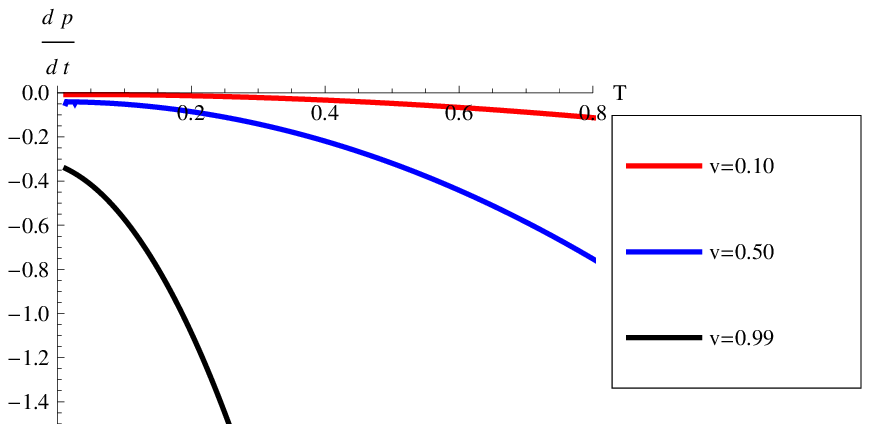} }} \caption{This figure shows the
 drag force as a function of $T$ for the chemical potential $\mu = 0.10$ (plot a), $0.34$ (plot b), and $0.80$ (plot c),
  respectively, in the hQCD model.  Here the dashed
curves stand for the behavior of drag force in confined phase which
is denoted by CF in the figure. In the confined phase, in fact the drag
force is not well defined,  meaning that the dashed curves do not make
any sense here.} \label{fig101}
\end{figure}

Certainly the analytic black hole solution (\ref{sol1}) does not
dual to a QCD model. As a warm-up exercise, we plot the drag force
in Figure [\ref{drag-T-mu ana}] for a model dual to this black hole
solution. We see that the drag force monotonically decreases with
temperature and for a fixed temperature it becomes large as chemical
potential grows. These features are qualitatively expected in
realistic QCD. However, the behaviors of jet quenching parameter and
screening length in the solution (\ref{sol1}) are far away from QCD
phenomenon and therefore we do not consider this solution from now
on.

\begin{figure}
\centering {\includegraphics[width= 10 cm]{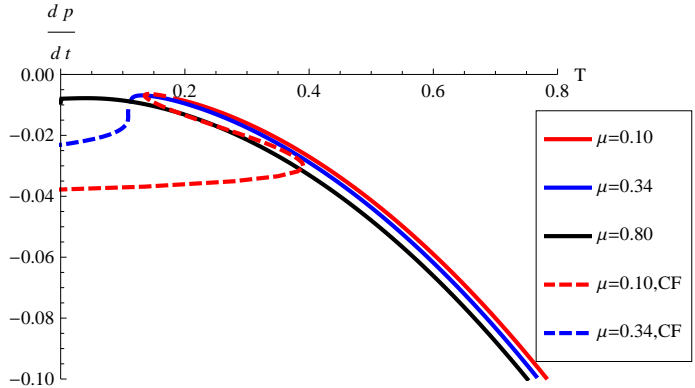} } \caption{The
figure shows the drag force as a function of $T$ for three chemical
potential $\mu =  0.10, 0.34$, and  $0.80$, respectively in the hQCD
model. The dashed parts of the curves stand for the drag force in
the confined phase which denoted by CF in the figure. Here we take
$v=0.1$.} \label{drag-3}
\end{figure}

In Figure  [\ref{fig101}] we plot the drag force for our hQCD model
given by the solution (\ref{sol2}) with different chemical
potentials $\mu=0.10$, $0.34$, and $0.80$, respectively. We can see
from the figure that for fixed chemical potential and temperature,
the drag force increases with the velocity of the quark, while for
fixed chemical potential and velocity, the drag force increases with
temperature. These are expected features in QCD. In particular, let
us note that in the low temperature region with small chemical
potential, the drag force is a multi-valued function of temperature
[see plot (a) and (b)], while it becomes a monotonic function with
large chemical potential [see plot (c)]. This feature is closely
related to the confinement/deconfinement phase transition in this
hQCD model \cite{Cai:2012xh}. The dashed parts of curves in plot (a)
and (b) denote the drag force in the confined phase and actually
they do not make any sense here since drag force is  not
well-defined in the confined phase.  Our result for the drag force
in the deconfined phase is in agreement with the one in
\cite{CasalderreySolana:2011us}.  For comparison, in Figure
[\ref{drag-3}] we plot the drag force versus temperature with three
different chemical potentials $\mu=0.10$, $0.34$ and $0.80$,
respectively. In this figure the velocity of quark is taken as
$v=0.1$.

\section{Jet Quenching parameter}

In this section, we use the AdS/CFT duality to compute the jet
quenching parameter in our hQCD model. The holographic method to
compute this quantity needs a consideration of Wilson loop
($\mathcal{C}$) traced out by a $q\bar{q}$ pair \cite{Liu:2006ug}.
The Wilson loop is taken to lie along the light cone in the gauge
theory. The gravity dual of this $q\bar{q}$ pair is represented as
the two end points of a fundamental string, attached to the boundary
of the bulk spacetime. Body of the string hangs along the radial
direction and up to the horizon of aAdS black hole. The Wilson loop
is mapped, in the dual theory, as the string world sheet. The jet
quenching parameter ($\hat{q}$) is related to the thermal
expectation value of the light-like Wilson loop operator, $\langle
{\mathcal{W}({\mathcal{C}_{light-like}})} \rangle$
\cite{Liu:2006he}. The holographic correspondence between thermal
expectation value of the light-like Wilson loop operator in
fundamental representation $\langle
{\mathcal{W}}^{F}(\mathcal{C}_{light-like}) \rangle$ and the
exponential of the worldsheet action, $e^{-S}$, leads us to obtain a
working formula of $\hat{q}$ in dual gravity theory. Here, $S$
stands for the worldsheet action of the fundamental string.
\begin{equation}
\langle{\mathcal{W}}^{F}(\mathcal{C}_{light-like}) \rangle =
\exp[-S(\mathcal{C})]. \label{adscft}
\end{equation}
In the planar limit, considering the fact $Tr_{Adj} = Tr^2_{Fund} $
, the relation between $W^F(\mathcal{C})$ and $W^A(\mathcal{C})$ can
be easily established as
 \begin{equation}
\langle W^A(\mathcal{C})\rangle = {\langle
W^F(\mathcal{C})\rangle}^2. \label{adfund}
\end{equation}
 Now we start with the background black hole solution in string frame
\begin{equation}
 ds_S^2=\frac{{\ell^2}
e^{2A_s}}{z^2}\left(-f(z)dt^2+\frac{dz^2}{f(z)}+dx^{1}dx^{1}+
dx^{2}dx^{2}+ dx^{3}dx^{3}\right). \label{jetmetric}
\end{equation}
By introducing the light cone coordinates  defined as
\begin{equation}
 x^{\pm} = \frac{t\pm x^1}{\sqrt{2}},
\end{equation}
the black hole metric (\ref {jetmetric}) can be rewritten as
\begin{eqnarray}
ds_S^2  &=&  \frac{{\ell^2}
e^{2A_s}}{z^2}\left(\frac{(1-f(z))}{2}({(dx^{+})}^2 + {(dx^{-})}^2)
- (1+f(z))(dx^{+}dx^{-}) \right. \nonumber \\
&& + \frac{dz^2}{f(z)}  + dx^{2}dx^{2}+ dx^{3}dx^{3}).
\end{eqnarray}
We take the gauge with  $\tau = x^- (0 \leq x^- \leq L^-)$, $\sigma
= x^2 (-\frac{L_2}{2} \leq x_2\leq \frac{L_2}{2})$, and
 set the pair of quarks at $x^2 = \pm \frac{L_2}{2}$ on $x^+ = $ constant,
$x^ 3 = $ constant plane. In the limit with $L^- \gg L_2$ the string
profile is completely specified by $z = z(\sigma)$.  Following
\cite{Liu:2006ug}, using (\ref {adscft}), (\ref {adfund}), one has
\begin{equation}
\langle W^A({\cal C})\rangle = \exp(-\frac{1}{4\sqrt{2}} \hat{q}
L^-L_2^2),
\end{equation}
where the jet quenching parameter is defined as
\begin{equation}
\hat{q} = \frac{8\sqrt{2}(S-S_0)}{L^{-}\ell^2},
 \label{formula}
\end{equation}
where $S$ is the Nambu-Goto action of the string and $S_0$ is the
self energy from  the mass of two quarks.

 Substituting the induced metric of the fundamental string
into the Nambu-Goto action (\ref {nambu-Goto}), we get
\begin{eqnarray}
S &=& - \frac{1}{2 \pi \alpha'} \int d\tau d\sigma \sqrt{-{\rm
det}g_{\alpha \beta}},  \nonumber \\
   &=&  \frac{L^- {\ell^2}}{\sqrt{2} \pi \alpha^{'}} \int^{\frac{L_2}{2}}_{0}  d\sigma
   \frac{e^{2A_s}}{z^2}\sqrt{(1-f(z))(1+\frac{{z^{'}}^2}{f(z)})}.
\label{jetaction}
\end{eqnarray}
Since the integrand in (\ref{jetaction}) does not explicitly depend
on $\sigma$, one can regard $\sigma$ as time and the integrand as a
Lagrangian. In this case the corresponding Hamiltonian is conserved.
That is, we can have
\begin{equation}
 \frac{\partial {\cal L}}{\partial z^{'}}z^{'} - {\cal L} = E,
\end{equation}
where $E$ is a constant and ${\cal L}$ is the integrand in
(\ref{jetaction}). From this relation we obtain the equation of
motion for $z$ as
\begin{equation}
{z'^{2}} = f(z) (\frac{e^{4A_s}}{z^4}\frac{(1-f(z))}{E^2} - 1).
\label{derevative}
\end{equation}

\begin{figure}
\centering {\includegraphics[width= 10 cm]{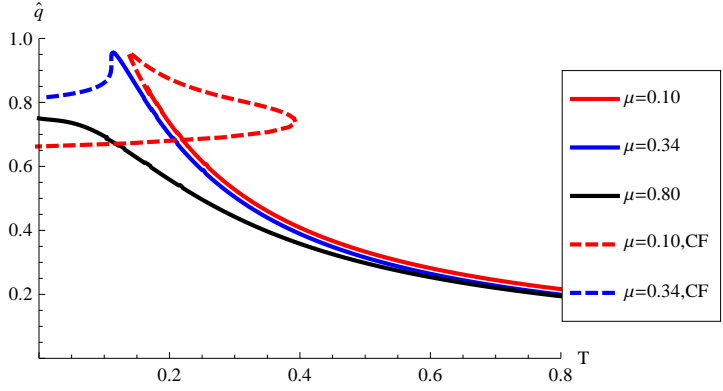} } \caption{The
figure shows the jet-quenching parameter as a function of $T$ for
three chemical potentials $\mu = 0.10, 0.34$, and $0.8$,
respectively, in the hQCD model. The dashed parts of curves stand
for the jet-quenching parameter in the confined phase which are denoted
by CF in the figure. } \label{qhat-T-mu}
\end{figure}

We choose the boundary conditions $z( \pm \frac{L_2}{2}) = 0$ and
$z^{'}(0) = 0$. In that case, the turning point $z_T$ is determined
by solving  Eq.(\ref {derevative}). Since $z^{'}(\sigma)$ is a real
function, so the square of it should be non-negative. The
realization of boundary condition $z^{'}(0) = 0$ at the turning
point requires the proper choices of zeros and the positivity region
of the right hand side of Eq.(\ref {derevative}). From the boundary
conditions of the black hole solution
\begin{equation}
 \displaystyle \lim_{z \to z_h} f(z) = 0 , \ \ \ \lim_{z \to 0} f(z) =
 1,
\end{equation}
together with the fact that we are interested in the case with small
$E$, it is clear that the factor \tiny $\frac{e^{4A_s}(1-f(z)) - E^2
z^4}{E^2 z^4}$ \normalsize is always positive near the black hole
horizon and negative near the boundary. To remove the region with a
negative ${z^{'2}}$, we consider a modified boundary at $ z = \delta
$. We assume that at $z = z_{min}$,
\begin{equation}
\frac{e^{4A_s(z_{min})}}{z_{min}^4}\frac{(1-f(z_{min}))}{E^2} -1 =
0,
\end{equation}
and $\delta > z_{min}$. In the region $\delta \leq z \leq z_h$,
thus,  the factor $[\frac{e^{4A_S}}{z^4}\frac{(1-f(z))}{E^2} -1]$ is
always positive. So only viable solution of $ {z^{'2}} = 0$ is
\begin{equation}
 f(z) = 0 \Rightarrow z_T = z_h.
\end{equation}
That is, the turning point is just at the horizon. The distance
between two quarks can be determined by
\begin{equation}
\frac{L_2}{2} = \int^{z_h}_{\delta} dz
\frac{E}{\sqrt{f[e^{4A_s}(1-f)z^{-4} - E^2]}}. \label{ac}
\end{equation}
As we are interested in the small $L_2$ limit,  considering the
smallness of $E$, we can expand Eq.(\ref {ac}) in terms of $E$ as
\begin{equation}
\frac{L_2}{2E} = \int^{z_h}_{\delta}dz
\frac{z^2 e^{-2A_s}}{\sqrt{ f(1-f)}} + \frac{E^2}{2}
 \int^{z_h}_{\delta}dz  \frac{e^{-6A_s}z^{6}}{\sqrt{f(1-f)^3}} +
 \mathcal{O}(E^4).
\label{acapp}
\end{equation}
Inverting (\ref {acapp}) suitably, we can obtain $E$ up to the
leading order of $L_2$ as
\begin{equation}
E= \frac{L_2}{2 \int^{z_h}_{\delta}dz
\frac{z^{2}e^{-2A_s}}{\sqrt{f(1-f)}}} +
\mathcal{O}({L_2}^3). \label{acappfin}
\end{equation}
Thus we can obtain the string action
\begin{eqnarray}
 S =   \frac{L^- {\ell^2}}{\sqrt{2} \pi \alpha^{'}}  \int^{z_h}_{\delta} dz
 \frac{e^{4A_s} (1-f)}{z^2\sqrt{f(e^{4A_s}(1-f)
 -z^4E^2)}}.
\label{jetaction2}
\end{eqnarray}
Clearly this action is divergent. The divergence comes from the
contribution of mass of two quarks. With the gauge $ x^- = \tau$ and
$ z = \sigma$,  the self energy of two free quarks reads
\begin{eqnarray}
S_0 =  \frac{L^- {\ell^2}}{\sqrt{2} \pi \alpha^{'}}
\int^{z_h}_{\delta} dz \frac{e^{2A_s}}{z^2}\sqrt{\frac{(1-f)}{f}}.
 \label{jetaction3}
\end{eqnarray}
Thus the regularized action up to the leading order of $L_2$ is
given by
\begin{equation}
 S_I = S - S_0 = \frac{L^-L_2^2\ell^2}{8\sqrt{2} \pi \alpha^{'}}\frac{1}{\int^{z_h}_{\delta}
  dz\frac{z^2e^{-2A_s}}{\sqrt{f(1-f)}}}+
\mathcal{O}({L_2}^4).
\end{equation}
With the definition of the  jet quenching parameter (\ref
{formula}), we finally reach
\begin{equation}
\hat{q} = \frac{\ell^2}{\pi \alpha^{'}}\frac{1}{\int^{z_h}_{\delta}dz
\frac{z^2e^{-2A_s}}{\sqrt{f(1-f)}}}.
\end{equation}
In fact the cutoff here can be removed  by noting the fact that the
integrand is regular inside the region $0 \leq z\leq z_h $, i.e,
from the horizon to the real boundary,
\begin{equation}
\int^{z_h}_{\delta}dz  \frac{z^2e^{-2A_s}}{\sqrt{f(1-f)}}
= \int^{z_h}_{0}dz  \frac{z^2e^{-2A_s}}{\sqrt{f(1-f)}}-
 \int^{\delta}_{0}dz  \frac{z^2e^{-2A_s}}{\sqrt{f(1-f)}}.
\end{equation}
The second integral in the right hand side of the above equation
smoothly vanishes in the  limit $\delta \to 0$. So the final
expression for the jet-quenching parameter is
\begin{equation}
\hat{q} = \frac{\ell^2}{\pi \alpha^{'}}\frac{1}{\int^{z_h}_{0}dz
\frac{z^2e^{-2A_s}}{\sqrt{f(1-f)}}}.
\end{equation}
Because the black hole metric is still too complicated to obtain an
analytical expression of the jet-quenching parameter in terms of
physical parameters,  we plot in Figure [\ref{qhat-T-mu}] the
jet-quenching parameter as a function of temperature in the hQCD
model with three chemical potentials $\mu=0.10$, $0.34$ and $0.80$,
respectively.  For large $\mu\geq \mu_c$ cases, the jet-quenching
parameter decreases monotonically with  temperature, which agrees
with the one in \cite{Gursoy:2009kk} qualitatively. On the other
hand,  when $\mu < \mu_c$, the jet-quenching parameter is a
multi-valued function of temperature in low temperature region and
it decreases monotonically with respect to temperature in high
temperature region. The multi-valued behavior of the jet-quenching
parameter in low temperature region is clearly related to the first
order phase transition between hadron phase (confined phase) and QGP
phase (deconfined phase). The jet-quenching parameter confirms the
hydrodynamical description of QGP phase and agrees with the real QCD
expectation in high temperature. Once again, as the drag force in
the confined phase, the dashed parts of curves in Figure
[\ref{qhat-T-mu}] denote the jet-quenching parameter in the confined
phase and thus they do not make any sense.

\section{Hot plasma wind and screening length}
The  screening length is defined as the maximum length achieved by a
quark-antiquark bound state at temperature $T > T_c$, beyond which
the pair dissociates. For quark-antiquark pair, the energetically
favorable configuration in the dual gravity theory is a fundamental
string with both ends attached to the boundary. The attached
endpoints correspond to the $q\bar{q}$ pair whereas being separated
beyond the screening length, thus dissociated from each other, the
pair maps into two separate strings hanging from the boundary. In
\cite{Fi: 2012de}, the screening length is computed in the rest
frame of $q\bar{q}$ pair and the plasma wind flows at a constant
speed $v$ for the hot $\mathcal{N} = 4$ SYM plasma. This setup is
identified with a quark-antiquark pair moving in hot $\mathcal{N} =
4$ SYM plasma.  In this section, we compute the screening length for
the hQCD model (\ref{sol2}) in the same way as in \cite{Fi: 2012de}.

In the static frame of $q\bar{q}$ pair, we assume that the hot
plasma is moving with velocity $v$ in the negative $x_3$ direction.
The Wilson loop we are interested in lies in the $t-x_1$ plane
specified by the length $\mathcal{T}$ and $L$ respectively. We
assume $\mathcal{T} \gg L$ such that the string world sheet is
invariant under translation along the time direction. The boost we
are considering is defined as
\begin{eqnarray}
dt &=& \cosh\eta dt^{'} -\sinh\eta dx^{'}_3 , \nonumber \\
 dx_3 &=&
-\sinh\eta dt^{'} + \cosh\eta  dx^{'}_3,
\end{eqnarray}
where  $\cosh\eta = \gamma$, $\sinh\eta = \gamma v$ and  $\gamma =
1/\sqrt{1-v^2}$ is the Lorentz boost factor. With the Lorentz
transformation, we obtain the boosted black hole metric in string
frame
\begin{eqnarray}
&ds_S^2  =& H(z) [-(1-(1-f)\cosh^2\eta)dt^2 +
(1+(1-f)\sinh^2\eta){(dx^{3})}^2,  \nonumber \\
&&- 2(1-f)\cosh\eta
\sinh\eta dt dx^3  + {(dx^{1})}^2+{(dx^{2})}^2+\frac{dz^2}{f(z)}],
\end{eqnarray}
where $H(z) = {\ell^2} e^{2A_s}/z^2$.
 We prefer to work in the static gauge
\begin{eqnarray}
\tau = t, \sigma = x^1, x^2(\sigma) = x^3(\sigma) = \text{constant},
\end{eqnarray}
with the following boundary conditions
\begin{eqnarray}
z(\sigma = \pm \frac{L}{2}) = 0, z(\sigma = 0) = z_c, z^{'}(\sigma =
0) = 0.
 \end{eqnarray}
Thus the world sheet metric induced on the boosted background is
given as
\begin{eqnarray}
g_{\tau \tau} &=& -H(z)(1-(1-f)\cosh^2\eta), \nonumber \\
g_{\tau \sigma} &=& g_{\sigma \tau} = 0, \nonumber \\
g_{\sigma \sigma} &=& H(z)[1 + (1  + \frac{{z^{'}}^2}{f(z)} )].
\label{sind}
\end{eqnarray}
Then the Nambu-Goto action for the string takes the form as
\begin{eqnarray}
 S = - \frac{\mathcal{T}}{\pi \alpha'} \int_{0}^{\frac{L}{2}} d\sigma  H(z) \sqrt{(1-(1-f)\cosh^2\eta )(1 +
 \frac{{z^{'}}^2}{f})}.
\label{sac}
\end{eqnarray}
As the Lagrangian ${\cal L}$ in (\ref{sac}) does not depend on
$\sigma $ explicitly, the corresponding Hamiltonian is conserved and
can be viewed as a constant of motion
\begin{eqnarray}
-q  =   \frac{\del\mathcal{L}}{\del {z}^{'}}{z}^{'} - \mathcal{L} .
\end{eqnarray}
With this we can cast the equation of motion in the form as
\begin{eqnarray}
z{'}  =   \frac{\sqrt{f[H^2(1-(1-f)\cosh^2\eta ) - q^2]}}{q}.
\label{scon}
\end{eqnarray}
It is evident from the constraint (\ref{scon}) that at the horizon,
$z = z_h$, where  $f(z_h)=0$, the factor
 $\frac{H^2}{q^2}(1-(1-f)\cosh^2\eta ) - 1 =  - \frac{H^2}{q^2} \sinh^2\eta -1$ is always negative. At
 the boundary, $f(0) = 1$, the factor $\frac{H^2}{q^2}(1-(1-f)\cosh^2\eta ) - 1
=  \frac{H^2}{q^2} -1$ is always positive for small values of $q
<H$.  Therefore in the range $0<z<z_h$ there must be a location
($z=z_c$) where
 $\frac{H^2}{q^2}(1-(1-f)\cosh^2\eta ) - 1$ switches its sign. Accordingly
  $z=z_c$ is the physical turning point of the string configuration. The
string can not be stretched up to the horizon as $z{'} $ is an
imaginary quantity in the region $z_c<z<z_h $. By solving the
equation
\begin{eqnarray}
\frac{f(z_c) H^2(z_c)\cosh^2\eta}{q^2} - \frac{H^2(z_c)\sinh^2\eta }{q^2} - 1 = 0,
\label{con}
\end{eqnarray}
the turning point can be numerically determined. Then one can obtain
the  binding energy between the quark and antiquark pair through
calculating the action (\ref{sac}) with constraint (\ref{scon})
\begin{eqnarray}
 V = -\frac{S-S_0}{\mathcal{T}},
\label{E}
\end{eqnarray}
where $S_0$ is given by
\begin{eqnarray}
S_0 = -\frac{\mathcal{T}}{\pi \alpha^{'}}\int^{z_c}_{0} dz
\sqrt{-G_{tt} G_{zz} }.
\end{eqnarray}
The distance between quark and antiquark can be calculated from
(\ref{scon}) as
\begin{eqnarray}
\frac{L}{2q} = \int^{z_c}_0 dz \frac{1}{ H
\sqrt{f[(1-\cosh^2\eta(1-f)) - \frac{q^2}{H^2}]}}. \label{L}
\end{eqnarray}
It is not possible to work out the integration in (\ref{L})
explicitly.  To determine the screening length,  we  plot the
distance $L$ with respect to the constant of motion $q$ for a fixed
rapidity $\eta$ in Figure [\ref{fig12}] (see plot (a)). It turns out
that for a fixed value of rapidity, there exists a maximum for $L$,
which is regarded as the screening length $L_s={L_{max}(\eta)/(\pi
T)}$. Plot (b) in Figure [\ref{fig12}] shows the  binding energy $V$
given by (\ref{E}) with respect to $L$.
\begin{figure}
\centering \mbox{\subfigure[]{\includegraphics[width=7
cm]{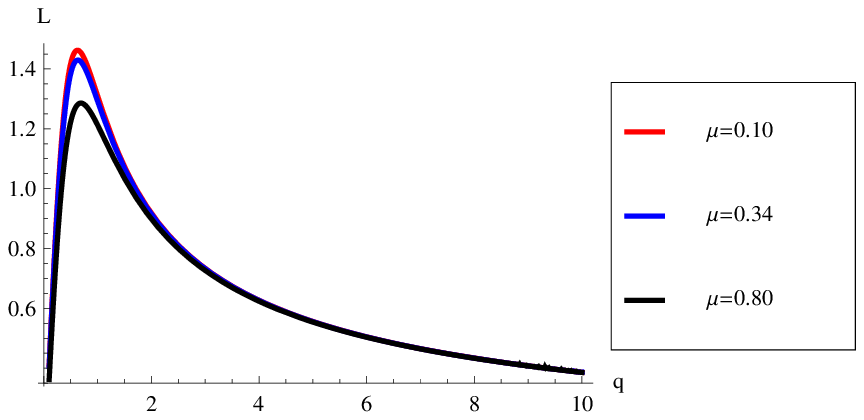}} \quad \subfigure[]{\includegraphics[width=7
cm]{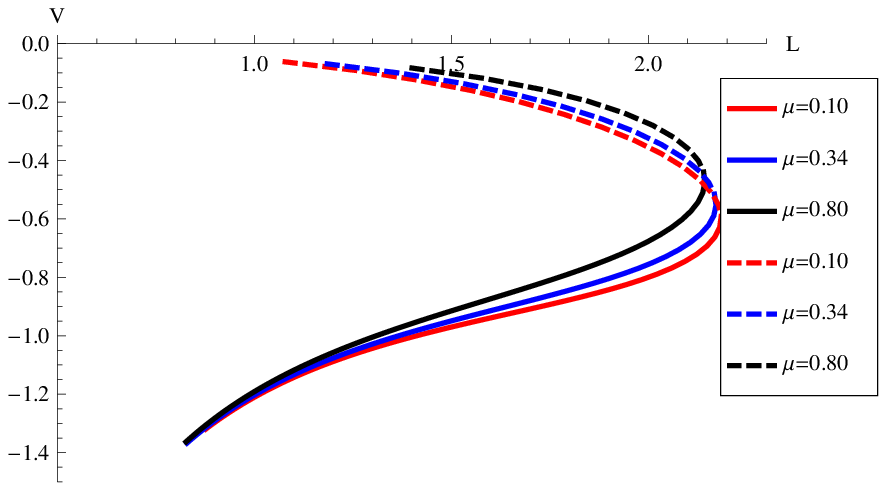} }} \caption{Plot (a) shows the quark-antiquark
distance as a function of $q$ for a fixed rapidity, while
 plot (b) shows the binding energy with respect to the distance.
  In both plots we fix the chemical potential $\mu =
0.10, 0.34$, and $0.80$, respectively. We have set a same
temperature $T$ to obtain these curves.} \label{fig12}
\end{figure}
One can see from plot (a) that the quark-antiquark distance starts
from zero when $q$ is also zero, it increases sharply with respect
to $q$, reaches its maximum at a certain $q$, and then decreases
monotonically to zero at some finite $q$. In between these two
zeros, there exists a single $L = L_{max}$ beyond which there is no
solution of Eq.(\ref{L}). This implies the quark-antiquark pair
dissociates beyond $ L = L_{max}$. We identify  $L_{max}(\eta)/(\pi
T)$ with the screening length $L_s$. For the $\mu=0.1$ case, $
L_{max}\simeq 1.4$ and $L_s\simeq {1.4/(\pi T)}\simeq{0.45/T}$,
close to the lattice calculation $L_s\sim {0.5/ T}$ \cite{O.
Kaczmarek} of the static potential between heavy quark and antiquark
in QCD. Plot (b) shows that there are two branches for the binding
energy in the region $L<L_{max}$. The branch with dashed curves has
a higher energy than the one with solid curves. This implies that
the branch with dashed curves is physically disfavored.

 \begin{figure}
 \centering {\includegraphics[width= 10 cm]{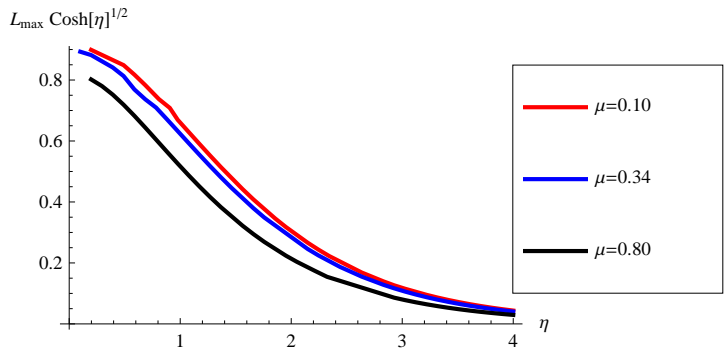}}
 \caption{The screening length versus the rapidity $\eta$ for the cases $\mu =0.1, 0.34$, and $0.8$,
 respectively. We have set a same temperature
$T$ to obtain these curves.}
 \label{screen-eta-mu}
 \end{figure}

The screening length $L_s(\eta)$ as a function of rapidity can be
obtained numerically as illustrated
 in Figure [\ref{screen-eta-mu}]. One finds that it decreases with
 velocity which indicates that the quark-antiquark pair dissociates at a
 lower temperature as it is moving. This behavior is also
observed in \cite{CasalderreySolana:2011us}. If the qualitative
behavior holds for QCD, it will have the consequence for quarkonium
suppression in heavy ion collision. Additionally, our results show
that the case with smaller chemical potential has a larger screening
length.

\section{Free energy and phase transition}

In this section, we would like to calculate the free energy of
the black hole solution dual to the hQCD model by following
\cite{Kiritsis:2012ma}.  Ref.~\cite{Kiritsis:2012ma} studies the
free energy of a generic Einstein-Maxwell-Dilaton system. The system
considered in \cite{Kiritsis:2012ma} is given by
\bea \label{introaa}
\begin{array}{rcl}
&&I=M_P^{d-1}\int_{M} d^{d+1}x \sqrt{-g} \left[ R-
\frac{Z(\Phi)}{4}F^2 -\frac{1}{2}(\partial \Phi)^2 +V(\Phi)
\right]+I_{GH}
~,\\[3mm]
&&I_{GH}=2 M_P^{d-1}\int_{\partial M} d^dx \sqrt{-h} K ~,
\end{array}
\eea with following ansatz
 \bea
  \label{introaf}
ds^2=e^{2\tilde{A}(u)}\left( -\tilde{f}(u)d t^2+d x^i d x^i\right)
+\frac{d u^2}{\tilde{f}(u)}~, ~~ {\bf A}=A_t(u)d t~, ~~ \Phi=\Phi(u)
~, \eea where $M_P$ is the Planck mass, and $K$ is the extrinsic
curvature of the finite boundary $\partial M$ with induced metric
$h$. By the following transformations
  \bea {d
\over d u}&=& e^{-\tilde{A}(u(z))} {d\over d z},{{\text{ }  \text{ }
\text{ } }}{d^2 \over d u^2}= e^{-2 \tilde{A}(u(z))}\Big({d^2\over d
z^2}-{d \tilde{A}(u(z))\over dz} {d\over d
z}\Big),{{\text{ } \text{ } \text{ } }} \tilde{f}(u(z))=f(z),\nonumber\\
\Phi(u)&=& \sqrt{{8\over 3}} \phi(z),{{\text{ }  \text{ } \text{ }
}} Z(\Phi)=1,{{\text{ }  \text{ } \text{ } }}  d=4,{{\text{ } \text{
} \text{ } }} M_P^{d-1}=\frac{1}{16 \pi G_5},\nonumber\\
\tilde{A}(u(z))&=&A_e(z)-\log(z),{{\text{ } \text{ } \text{ } }}
-V(u(z))=V_E(z), \label{relationbetween} \eea
 we can change the system and ansatz to ours discussed in this
 paper. Here we have chosen $\tilde{A}(u(z))=A_e(z)-\log(z)$ as a gauge. Varying the
action (\ref{introaa}) yields equations of motion for the
gravitational field
\begin{eqnarray}
\label{EOMan}
E_{\mu\nu}+\frac{1}{2}g_{\mu\nu}\left(\frac{1}{2}
\partial_{\mu}\Phi\partial^{\mu}\Phi-V(\Phi)\right)
-\frac{1}{2}\partial_{\mu}\Phi\partial_{\nu}\Phi -\frac{Z(\Phi)}{2
}\left(F_{\mu \rho}F^{\rho}_{\nu}-\frac{1}{4}
g_{\mu\nu}F_{\rho\sigma}F^{\rho\sigma}\right)=0.
\end{eqnarray}
The Maxwell equation and the explicit forms of Einstein equations
read
\begin{subequations}
\bea \label{introaia} \frac{d}{d u}\left( e^{(d-2)\tilde{A}}Z \dot
A_t\right)&=&0~,\\ \label{introaib} 2(d-1)\ddot {\tilde{A}}+{\dot
\Phi}^2&=&0~,
\\\label{introaic} \ddot {\tilde{f}}+d \dot {\tilde{A}} \dot {\tilde{f}}-e^{-2\tilde{A}} Z {\dot A_t}^2&=&0~,
\\ \label{introaid} (d-1)\dot {\tilde{A}} \dot {\tilde{f}}+\left(d(d-1) {\dot
{\tilde{A}}}^2-\frac{1}{2} {\dot \Phi}^2 \right){\tilde{f}}-V
+\frac{1}{2} Ze^{-2\tilde{A}} {\dot A_t}^2&=&0 ~, \eea
\end{subequations}
where an over dot stands for the derivative with respect to $u$.
Eq.(\ref{introaia}) is the Maxwell equation and the other three
Eq.(\ref{introaib}) (\ref{introaic}) (\ref{introaid}) are obtained
from Einstein equations. One can easily check that
Eq.(\ref{introaib}) and  Eq.(\ref{introaic}) correspond to
Eq.(\ref{AF}) and Eq.(\ref{ff}), respectively. In addition, the
equation of motion for the scalar field is \bea {1\over
\sqrt{-g}}\partial_u
\left(\sqrt{-g}\partial_u\Phi\right)+V'(\Phi)-{Z'(\Phi)\over
4}F^2=0, \eea where the prime denotes the derivative with respect to
$\Phi$. From (\ref{introaia}), one  obtains  \bea
A_t(u)=\mu+\int_{u_0}^u {\rho\over
e^{(d-2)\tilde{A}(\tilde{u})}Z(\Phi)} d\tilde{u}, \eea  where $\mu$
and $\rho$ are the chemical potential and charge density of the
black hole solution, respectively, and $u_0$ stands for the UV
boundary.

By defining the superpotential $W$ in the following way
\cite{Kiritsis:2012ma}
 \bea \label{introal} \dot \Phi=W'(\Phi),
  \eea
the equation \eqref{introaib} can be solved as
 \bea \label{introam}
\dot {\tilde{A}}=-\frac{W(\Phi)}{2(d-1)} ~. \eea Equivalently, \bea
\label{introan}
\tilde{A}(\Phi)=\tilde{A}_0-\frac{1}{2(d-1)}\int_{\Phi_0}^\Phi
\dd\tilde \Phi\, \frac{W(\tilde\Phi)}{W'(\tilde\Phi)}~, ~~
u=u_0+\int_{\Phi_0}^\Phi \frac{\dd\tilde \Phi}{W'(\tilde\Phi)}, \eea
where $\tilde{A}_0=\tilde{A}(\Phi_0)$, $\Phi(u=u_0)=\Phi_0$.

The temperature $\tilde{T}$ associated with the black hole in metric
(\ref{introaf}) is given by
 \bea \label{introaq}
\tilde T= \frac{1}{4\pi} \left|e^{\tilde{A}} \dot {\tilde{f}
}\right|_{u=u_h}, \eea
 where $u_h$ denotes the black hole horizon. The entropy density
 for the black hole solution (\ref{introaf}) is
 \bea \label{introas}
\tilde{S}= 4\pi  e^{(d-1)\tilde{A}(\Phi_h)}= 4\pi
e^{(d-1)\tilde{A}_0-\frac{1}{2}\int_{\Phi_0}^{\Phi_h}\dd\tilde \Phi
\frac{W}{W'}}, ~~~ (\Phi_h:=\Phi(u_h)). \eea
 where $M_P^{d-1}=1$ has been taken.

The on-shell action $I$ has a bulk contribution $I_E$ and a boundary
contribution $I_{GH}$ from the Gibbons-Hawking term. Here we
evaluate it with an UV cutoff at $u=u_0$ or $z=\epsilon$.
 A standard
computation gives us with the action density
  \bea \label{effpotbd}
I_{on-shell}&=&\tilde{T}^{-1}  (-W+\dot {\tilde{f}} )e^{d
\tilde{A}(u_0)}|_{u=u_0}.\eea
 With the equations
(\ref{relationbetween})(\ref{introaic}) and $A_t(u_h)=0$, the free
energy density is found to be
  \bea \label{effpotbf} F&=&
-e^{d\tilde{A}_0} W(\Phi_0) -  \tilde{T} \tilde{S} +\rho^2
\int_{\Phi_h}^{\Phi_0}\frac{d \tilde \Phi}{e^{(d-2)\tilde{A}}Z(\Phi)W'(\Phi)},\nonumber\\
&=&{6 b^{3}({z}) {d\over d z}\Big(A_e(z)-\log
(z)\Big)\Big|_{z=\epsilon}- T S +{\mu^2 \over
\int_{z_h}^{\epsilon}\frac{dz}{b(z)Z(\phi)}}~}, \label{effpotbfinal}
\eea
 where $ b(z)=e^{A_e(z)}/z$. Note that in the second line of
 Eq.(\ref{effpotbfinal}), we have considered the case $d=4$ and the
 expression of the free energy is written in the ansatz
 (\ref{metric-Einsteinframe}) in Einstein frame.
The first term in Eq.(\ref{effpotbfinal}) corresponds to the black
hole mass and the last term can be expressed as $-\mu\rho$.

\begin{figure}
\centering {\includegraphics[width= 9 cm]{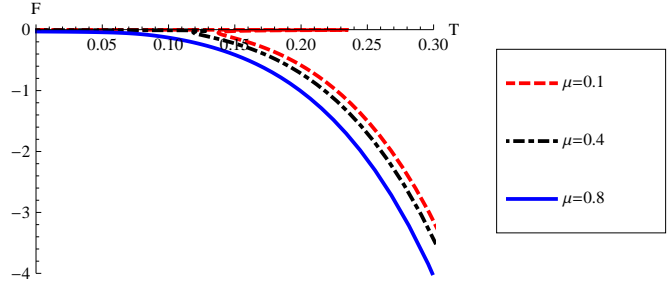}}
\caption{The free energy of the background black hole solution
with respect to temperature.  It shows that in the small $\mu <\mu_c$ case,
the free energy has a multi-valued behavior, while it is absent
in the large $\mu\geq\mu_c$ case. The parameters $\ell=1,16 \pi G_5=1, g_g=1, k=0.3
\text{GeV}$ are taken here.}
 \label{freeenergy-t}
\end{figure}

With the analytical result (\ref{effpotbfinal}), one can obtain the
free energy of the black hole solution (\ref{sol2}) with respect to
temperature with  various chemical potentials numerically. Figure
[\ref{freeenergy-t}] shows the free energy density of the black hole
solution. It can be seen that the free energy density is a
multi-valued function of temperature in the case $\mu< \mu_c$ and it
becomes monotonically  in the large $\mu$ region $\mu>\mu_c$. The
multi-value behavior of the free energy indicates the existence of a
first order phase transition in the small $\mu$ region.

 To clearly see this, it should be helpful to recall the
temperature behavior of the black hole (see Figure [\ref{T-zh}]).
One can see that when $\mu >\mu_c$, the black hole is always stable
with positive heat capacity. When $\mu <\mu_c$, one can see from
plot (b) that as $T>T_M$, the black hole is also always stable with
positive heat capacity (in branch 1), when $0<T<T_m$, the black hole
is also stable with positive heat capacity (in branch 3), while when
$T_m <T<T_M$, there exist three black hole solutions with a same
temperature. For instance, for temperature $T_p$, there exist three
black hole solutions with horizon radii $z'_p$, $z_p$ and $z''_p$,
 in branch 1, 2 and 3, respectively. The black hole solutions in branch 2 are
 unstable because those black holes have negative heat capacity. We call black holes in
 branch 1 and branch 3 as large and small black holes, respectively, when $T_m <T<T_M$,
 because the black holes in branch 1 have  large horizon radius and entropy, while
 the black holes in branch 3 have  small horizon radius and entropy.
 From Figure [\ref{T-zh},\ref{freeenergy-t}], we can see that when $\mu >\mu_c$, the black hole solution
 always dominates, which indicates the dual field theory is in deconfined phase, while in the
 $\mu < \mu_c$ regime, when $T_m <T <T_M$, there exists a phase transition between large
 black hole and small black hole. In this transition, the entropy has a jump, which means
 this is a first order phase transition. On the other hand, when $0<T<T_m$, the small black hole
 solution in branch 3 is dominated. This is quite different from the case of Schwarzschild black hole
 in AdS space. There when temperature is less than a certain value, there does not exist
 black hole solution. In this case, the thermal gas is dominated. Above a certain temperature,
 there exist two black hole solutions, one is stable and the other is unstable. Therefore
 there exists a well-known first order phase transition (Hawking-Page phase transition) between
 black hole and thermal gas in AdS space~\cite{Hawking-Page}. This transition is explained as the confinement/deconfinement
 phase transition of dual field theory. In our case even in the low temperature regime with $T<T_m$, the
 small black hole is dominated, the free energy for these small black holes is negative. Compared
 to thermal gas, those black hole solutions are still favored. When temperature increases beyond $T_m$,
 the black hole solution with large horizon radius in branch 1 is favored than the one with small horizon
 radius in branch 3. In this case, a phase transition happens. Since the entropy of small black holes in
 branch 3 is much small than the one of large black holes in branch 1, this phase
 transition therefore
 can be identified as the confinement/deconfinement phase transition in the dual theory.
 Of course, this phase transition is not same as the Hawking-Page phase transition. The free energy calculation
 further confirms this picture. And this picture is completely consistent with
 the one in \cite{Cai:2012xh} by computing heavy quark potential in this model. In addition,
 we would like to stress that the phase transition between large black hole and small black hole
 is the same as the one found in \cite{Chamblin:1999tk}, but the transition in our case is in grand
 canonical ensemble, while the one in \cite{Chamblin:1999tk} is in canonical ensemble. As a result,
 $T-Q$ phase diagram in Figure 1 of \cite{Chamblin:1999tk} is similar to our $T-\mu$ phase diagram.

 In a word, in the small $\mu$ region there exists a first order
phase transition, while it is absent in the large $\mu$ region
 through studying the free energy of the system. The
existence of the first order phase transition in the small $\mu$
region is consistent with the analysis of Wilson loop in this black
hole background~\cite{Cai:2012xh}. The existence of the critical
point at $\mu=\mu_c$ is in agreement with recent lattice calculation
given in \cite{Fromm:2011qi}. Let us notice that in
\cite{Cai:2012xh} we claim that there is a crossover or higher order
phase transition in the large $\mu>\mu_c$ region, while the free
energy calculation here shows no such phase transition. In this
sense, free energy may show less information about phase transition
and one should take more probes into considerations, such as
Polyakov loop \cite{Cai:2012xh}, to understand the phase structure
completely. These two conclusions are in fact not in contradiction
with each other. The reason is as follows. The free energy
calculation here only concerns with the black hole background, from
the point of view of QCD, the black hole background does not include
the degrees of freedom of quarks, while the Wilson loop calculation
in \cite{Cai:2012xh} is related to the dynamics of quarks.

 In Figure [\ref{Phasedigram-two}] we plot the phase diagram of the
 hQCD model. The difference between the confinement/deconfiement phase transition
 lines come from the different consideration as mentioned above.

\begin{figure}
\centering {\includegraphics[width= 7 cm]{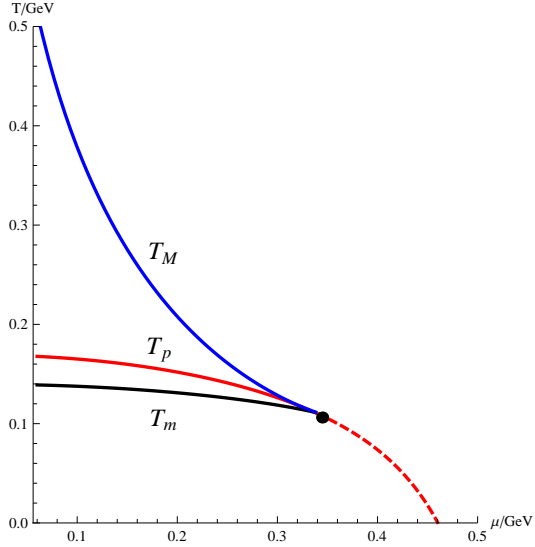}
\label{Phasedigram-two}}\caption{The phase diagram of the hQCD
model. The red and dashed red curves are obtained in
\cite{Cai:2012xh} through studying VEV of Polyakov loop. The two
curves stand for first order and continuous phase (or second order)
transition respectively.  The thick black and blue curves correspond
to the minimal temperature $T_m$ and maximal temperature $T_M$ in
small $\mu$ region.  The thick black curve is obtained by
calculating the free energy. The black dot denotes the critical
point.  $T_p$ denotes the transition
temperature.}\label{Phasedigram-two}
\end{figure}

At the end of this section, let us stress the question whether
 there exists the so-called Hawking-Page phase transition  between black hole and thermal gas
 in our model.  Note that in the above analysis, our
phase transition happens between small black hole and large black
hole in the region with small chemical potential $\mu <\mu_c$, such
a transition does not occur as $\mu >\mu_c$.  It is therefore of
some interest to investigate whether there exists any
 transition between black hole solution and thermal gas solution in our hQCD model.
For this aim, we have to first find the thermal gas solution. Unfortunately, in this
 potential reconstruct approach, to find out the thermal gas solution is not an easy job
 since we have not an explicitly analytic form of the potential for the
 dilaton field in our hQCD model.  But even so, we still can make some arguments which
  lead to the conclusion that there does not exist any phase transition between
  black hole and thermal gas solutions in the hQCD model. The main reason for this comes
  from the fact that in the hQCD model, for any temperature, there exist corresponding
  black hole solutions.  This can be seen from Figure [\ref{T-zh}]. It shows in plot (a)
  that when $\mu >\mu_c$, the black hole temperature increases monotonically from zero with the smallest
  black hole radius, which corresponds to an extremal black hole, while when $\mu <\mu_c$, plot (b) shows
  that as $T \le T_m$, there exists a region with stable black holes with small horizon radius, including the
  extremal black hole; and as $T>T_m$, there exists the region with stable black holes with larger horizon
  radius. In addition, for a given temperature, it is generally believed that the black hole solution has much
  large entropy than the thermal gas, so that the black hole phase is always dominated. As a result,
  whatever the chemical potential is, the black hole solutions are always dominated over the thermal
  gas solutions. This is quite different from the case of Schwarzschild-AdS black hole. In the latter case,
  there exists a minimal temperature, below which there is no black hole solution. This leads to the
  conclusion that in low temperature the thermal gas solution is favored, while in high temperature
  the black hole solution is dominated, and thus there must exist the Hawking-Page phase transition
  between the black hole solution and thermal gas solution at a certain temperature. Therefore in our
  case, black hole solutions are always dominated and there does not exist any phase transition
  between black hole solution and thermal gas solution. This might be a common feature for black hole solutions
  with Ricci flat horizon in AdS space.  But as we analyzed above, there exists indeed
  the phase transition between small black hole and large black hole in our model with $\mu <\mu_c$.
  This phase transition can also be understood as the confinement/deconfinement phase transition in QCD
  since clearly the small black hole has a much less entropy than the large
  one. And this interpretation is also consistent with the
  calculation of heavy quark potential in \cite{Cai:2012xh}.

\section{Conclusion and discussion}

In this paper, we have continued to study the holographic QCD model
proposed in \cite{Cai:2012xh}, in an Einstein-Maxwell-Dilaton
system.  At first we have generalized case with a non-minimal
coupling between Maxwell field and dilaton field, and given a
generic formulism for generating a set of exact and asymptotic AdS
black hole solutions in the EMD system. After briefly reviewing the
main features of the hQCD model, we have studied some aspects of QGP
phase of the hQCD model by calculating some quantities such as drag
force, jet quenching parameter and screening length. The
calculations show that the behaviors of those quantities are
consistent with the expectation from realistic QCD.

 It is found that the drag force increases
monotonically with temperature which is quite good consistent with
real QCD phenomenon in the larger chemical potential region with
 $\mu\geq \mu_c$. In the small chemical region with $\mu <\mu_c$,
 the drag force  also monotonically increases in the high temperature region, while
 in the low temperature, it shows a multi-valued behavior.  Note that
 in the case $\mu<\mu_c$, the solution is dual to the confined phase of QCD.
 In that case, the drag force is not well defined. Therefore the change from the
 multi-valued behavior to the monotonic behavior just manifests the existence of
 the first order phase transition.  The jet quenching parameter has
monotonically decreasing behavior versus temperature, which is also
consistent with QCD experiments in $\mu\geq \mu_c$. For the $\mu<\mu_c$
case,  the jet quenching parameter agrees with real QCD expectation in
high temperature and once again, it shows the multi-valued behavior in
the low-temperature region. As in the case of drag force, the multi-valued
behavior of jet quenching parameter in the low temperature region is
consistent with the existence of first order phase transition in this hQCD
model.  For the screening length
we have plotted  the separation between quark and anti-quark with respect to
the constant of motion $q$. It is clear from the plot that for both
cases when  $\mu\geq \mu_c$ and $\mu<\mu_c$ the dipole dissociates
beyond a maximum separation distance, namely the
screening length $L_s$. We have also calculated the binding energy as a
function of separation distance. In addition, we have presented $L_s(\eta)$ and
found that there are qualitative consequences for quarkonium
suppression in heavy ion collisions in this hQCD model.

We have calculated the free energy of the background
 black hole solution and further confirmed that there exists a first order
 phase transition in small $\mu$ region between small black hole and
large black hole~\cite{Cai:2012xh}. When $\mu >\mu_c$, the phase
transition is absent from the point of view of the free energy. The
existence of the critical point is consistent with the result
in~\cite{Cai:2012xh}. Further we have argued that there does not
exist any phase transition between black hole solutions and thermal
gas solutions in this model, and the main reason is given.

In this work we have studied some aspects of QGP phase in the hQCD
model proposed in~\cite{Cai:2012xh}. The results are encouraged and
are consistent with the expectation of real QCD.  Thus it would be
of great interest to further investigate the hQCD model. For
example, it is required in the model to study the spectra of
hadrons, chiral phase transition and its phase diagram
\cite{Evans:2011mu}\cite{Evans:2011tk}\cite{Evans:2011eu},
hydrodynamical properties of QGP, and color flavor locked phase
\cite{arXiv:0909.1296}\cite{arXiv:1101.4042}\cite{hep-ph/0302142},
etc. Furthermore, quantum corrections \cite{Singh:2012xj} to above
physical quantities are also deserved to consider in the coming
works.

 Finally we would like to mention that in this paper the drag
force, jet quenching and screening length have been calculated
through the Nambu-Goto action of fundamental string in the AdS/CFT
correspondence. As one knows that the Nambu-Goto action makes sense
only in string theory. Therefore it becomes a crucial issue whether
our hQCD model can be embedded into some string theory. This is also
a key point for a large kind of phenomenal models of holographic
QCD. At the moment, we cannot show that the model could be a
consistent truncation of some low energy effective theory of string
theory, but we hope this model is helpful to build a holographic
 model dual to the realistic QCD.

\section*{Acknowledgements}

The authors are grateful to Jiunn-Wei Chen, Xiaofang Chen, Danning
Li, Sudipta Mukherji, Ajit M.Srivastava, Shi Pu, Wen-Yu Wen,
Shang-Yu Wu, Chen-Pin Yeh, Yi Yang and Yun-Long Zhang for useful
discussions. This work was supported in part by the National Natural
Science Foundation of China (No.10821504, No.10975168 and
No.11035008), and in part by the Ministry of Science and Technology
of China under Grant No. 2010CB833004. SC wishes to thank the
members of ITP, CAS for the warm hospitality during initial part of
this work. SH would like to thank the Department of Electrophysics
of National Chiao-Tung University, National Taiwan university and
the ``International School On Strings And Fundamental Physics" held
at Hamburg, for their hospitality and financial support. SH also
would like to appreciate the general financial support from China
Postdoctoral Science Foundation with No. 2012M510562.

\end{document}